\documentclass[
  shortnames,nojss]{jss}

\usepackage{orcidlink,thumbpdf,lmodern}

\usepackage[utf8]{inputenc}

\author{
Louis J. M. Aslett~\orcidlink{0000-0003-2211-233X}\\Durham
University \And Ryan R. Christ~\orcidlink{0000-0002-2049-3389}\\Yale
University
}
\title{\pkg{kalis}: A Modern Implementation of the Li \& Stephens Model
for Local Ancestry Inference in \proglang{R}}

\Plainauthor{Louis J. M. Aslett, Ryan R. Christ}
\Plaintitle{kalis: A Modern Implementation of the Li \& Stephens Model
for Local Ancestry Inference in R}
\Shorttitle{\pkg{kalis}: Local Ancestry Inference}

\Abstract{
Approximating the recent phylogeny of \(N\) phased haplotypes at a set
of variants along the genome is a core problem in modern population
genomics and central to performing genome-wide screens for association,
selection, introgression, and other signals. The Li \& Stephens (LS)
model provides a simple yet powerful hidden Markov model for inferring
the recent ancestry at a given variant, represented as an \(N \times N\)
distance matrix based on posterior decodings. However, existing
posterior decoding implementations for the LS model cannot scale to
modern datasets with tens or hundreds of thousands of genomes. This work
focuses on providing a high-performance engine to compute the LS model,
enabling users to rapidly develop a range of variant-specific ancestral
inference pipelines on top, exposed via an easy to use package,
\pkg{kalis}, in the statistical programming language \proglang{R}.
\pkg{kalis} exploits both multi-core parallelism and modern CPU vector
instruction sets to enable scaling to problem sizes that would
previously have been prohibitively slow to work with. The resulting
distance matrices enable local ancestry, selection, and association
studies in modern large scale genomic datasets.
}

\Keywords{\proglang{R} package}
\Plainkeywords{R package}

\Address{
    Louis J. M. Aslett\\
    Durham University\\
    Department of Mathematical Sciences,\\
Stockton Road,\\
Durham,\\
DH1 3LE,\\
United Kingdom.\\
  E-mail: \email{louis.aslett@durham.ac.uk}\\
  URL: \url{https://www.louisaslett.com/}\\~\\
      Ryan R. Christ\\
    Yale University\\
    School of Medicine\\
333 Cedar Street,\\
New Haven,\\
CT 06510,\\
United States of America.\\
  E-mail: \email{ryan.christ@yale.edu}\\
  
  }

\providecommand{\tightlist}{%
  \setlength{\itemsep}{0pt}\setlength{\parskip}{0pt}}

\usepackage{amsmath}
\usepackage{booktabs}
\usepackage{caption}
\usepackage{longtable}

\usepackage{amsmath} \usepackage{algorithm} \usepackage{algpseudocode} \usepackage{amsfonts} \usepackage{cleveref} \usepackage{fvextra} \DefineVerbatimEnvironment{CodeOutput}{Verbatim}{breaklines,commandchars=\\\{\}}

\begin{document}

\hypertarget{introduction}{%
\section{Introduction}\label{introduction}}

The hidden Markov model (HMM) of haplotype diversity proposed in
\citet{Li2213} (hereinafter, the LS model) has become the basis for
several probabilistic phasing, ancestry inference, and demographic
inference methods in modern genomics \citep[\citet{speidel}]{Song1005}.
The LS model provides the basis for the genome-wide ancestry inference
software ChromoPainter, which summarizes the ancestry of \(N\)
haplotypes with an \(N \times N\) similarity matrix
\citep{lawson2012inference}. This matrix is obtained by running \(N\)
independent HMMs in which each haplotype is modelled as a mosaic of all
of the other haplotypes in the sample. This \emph{`all-vs-all'} copying
approach is motivated by the product of approximate conditionals (PAC)
likelihood originally proposed by \citet{Li2213} and allows
ChromoPainter to render a chromosome-wide estimate of the recent
ancestry of the \(N\) haplotypes with high resolution.

Beyond chromosome-wide summaries, the LS model can also be used for
variant-specific ancestry inference. The contemporary importance of this
is highlighted by the recent advances achieved in the RELATE software
suite \citep{speidel}, which uses the LS model internally to initialise
variant-specific ancestral trees for downstream population genetic
analyses ranging from demography to selection inference. \pkg{kalis}
focuses on providing a high-performance engine to compute exclusively
the LS model, enabling users to rapidly develop a range of future
variant-specific ancestral inference pipelines on top, in the easy to
use statistical programming language \proglang{R}.

At the same time, it has been recognised for over a decade
\citep{sutter2005free} that the serial execution speed of CPUs will
increase modestly, with additional performance primarily coming from
concurrency via multi-core architectures or the growing width of
specialised single instruction, multiple data (SIMD) instruction sets.
Whilst multi-core architectures are now somewhat routinely exploited via
forked processes or threading, SIMD instructions remain an often
overlooked source of performance gains, possibly because they are harder
to program. There are a cornucopia of SIMD instruction sets: on the
Intel platform the genesis was in the 64-bit wide MMX instruction set
\citep{mmx} which allows simultaneous operation on two 32-bit, four
16-bit or eight 8-bit integers. The most recent incarnation on Intel
CPUs is a suite of AVX-512 instruction sets \citep{intelisa}, now
capable of operating on 512-bits of different data types simultaneously
(eg eight 64-bit floating point, or sixteen 32-bit integer values).
Other CPU designs have similar SIMD technologies, such as NEON on ARM
CPU \citep{armneon} designs (including the Apple M1 and M2 processors,
as well as Amazon Web Services Graviton range). Additionally all modern
CPUs are superscalar architectures supporting instruction level
parallelism, an advance that has been in the consumer Intel platform
since the Pentium \citep{pentium}. Judicious programming can make it
easier for compilers and the deep reorder buffers of modern pipelined
CPUs to exploit this more hidden form of parallelism.

In this work we provide a reformulation of the LS model and an optimised
memory representation for haplotypes, which together enable us to
leverage \emph{both} multi-core and SIMD vector instruction parallelism
to obtain local genetic distance matrices for problem sizes that
previously appeared out of reach. This high performance implementation
is programmed in \proglang{C} \citep{C18}, with an easy to use interface
provided in \proglang{R} \citep{R}. We provide low-level targets of
AVX2, AVX-512 and NEON instruction sets (covering the vast majority of
CPUs in use today), and the whole package has an extensive suite of
\(162,835\) unit tests.

In \Cref{sec:algorithm} we describe the background of the LS model and
our reformulation which makes it amenable to these high-performance CPU
technologies. In \Cref{sec:thepkg} we describe the user friendly
\proglang{R} interface which enables easy use of the high performance
implementation without any knowledge of the underlying CPU technologies,
whilst in \Cref{sec:core} we describe the technical details of the
underlying low-level implementation for the interested reader (note:
\pkg{kalis} can be fully utilised without reading \Cref{sec:core}).
\Cref{sec:perf} demonstrates the performance that can be achieved with
\pkg{kalis}, including examples with 100,000 haplotypes. To the best of
our knowledge, this is the first example of running the LS model at the
scale of hundreds-of-thousands of haplotypes. Finally, in
\Cref{sec:realdata} we present a real data example using \pkg{kalis} to
examine the ancestry at the \emph{LCT} gene, and present future work in
\Cref{sec:future}.

\hypertarget{sec:algorithm}{%
\section{Algorithm}\label{sec:algorithm}}

We recap the LS model and fix our notation in \Cref{sec:lsmodel}, after
which we will present the reformulation which enables high performance
computation in \Cref{sec:reformulation}.

\newpage

\hypertarget{sec:lsmodel}{%
\subsection{The LS model}\label{sec:lsmodel}}

To formalize our objective, let \(h\) be an \(L \times N\) matrix of
\(0\)s and \(1\)s encoding \(N\) phased haplotypes at \(L\) sites. Let
\(h_i^{\ell} \in \{0,1\}\) denote the the \((\ell,i)\)th element of
\(h\). For brevity, let \(h_i\) denote the \(i\)th haplotype (the
\(i\)th column of \(h\)) and \(h_{-i}\) denote all of the haplotypes
excluding the \(i\)th haplotype. The LS model proposes an HMM for
\(h_i | h_{-i}\) in which the hidden state at variant \(\ell\),
\(X^{\ell}_i \in \{1,\dots,N\} \setminus i\), is an index indicating the
haplotype in \(h_{-i}\) that \(h_i\) is most closely related to (or
``copies from'') at variant \(l\). We present here their proposed
emission and transition kernels (see Equation A1 and Equation A2 in
\citet{Li2213}) with a simplified parametrisation that is similar, but
not identical, to that used by ChromoPainter.

While the original LS model assumes that each haplotype has an equal
\emph{a priori} probability of copying from any other, following
ChromoPainter, we define a left stochastic matrix of prior copying
probabilities \(\Pi \in \mathbb{R}^{N \times N}\) where \(\Pi_{ji}\) is
the prior probability that haplotype \(j\) is copied by \(i\) and, by
convention, \(\Pi_{ii} = 0\). Here and whenever possible in \pkg{kalis},
all matrices are column-oriented such that the \(i\)th column pertains
to an independent HMM where \(h_i\) is treated as the observation. There
is some probability of a mis-copy at variant \(\ell\), \(\mu^{\ell}\),
which under the LS model is set proportional to the mutation rate at
\(\ell\). This leads to an emission kernel of the form \begin{equation}
  \theta_{ji}^{\ell} := \mathbb{P}\left(h_{i}^\ell \left| X_{i}^{\ell} = j \right. \right)
  = \begin{cases}
    1 - \mu^{\ell} & \text{if } h_{i}^\ell = h_j^\ell \\
    \mu^{\ell} & \text{if } h_{i}^\ell \neq h_j^\ell \\
  \end{cases} .
  \label{eq:emission}
\end{equation}

The transition kernel between hidden states is based on the
recombination rate between sites. Let \(m^l\) be the genetic distance
between variant \(l\) and variant \(l+1\) in Morgans (the expected
number of recombination events per meiosis). Define
\(N_e = 4\tilde{N_e}/N\) where \(\tilde{N_e}\) is the effective diploid
population size (ie half of the haploid effective population size).
Then, under the LS model the transition kernel is \begin{equation}
  P(X_{i}^\ell = k | X_{i}^{\ell-1} = j)
  = \Pi_{ki} \rho^\ell + \mathbf{1}\left\{k = j\right\} \left(1-\rho^\ell\right) ,
  \label{eq:transition}
\end{equation} where \(\rho^\ell = 1-\exp\left(-N_e m^\ell\right)\) and
\(\mathbf{1}\left\{\cdot\right\}\) is the indicator function.
\citet{Li2213} observe that in practice the estimation of recombination
rates is improved when the scaled recombination rate is raised to a
power, so we adopt this approach and introduce an exponent \(\gamma\).
For \(\gamma>1\) the recombination map becomes more heavily peaked,
whereas \(\gamma<1\) tempers the recombination map to make it more flat
and smooth. Hence, in \pkg{kalis}, we set \begin{equation}
  \rho^\ell := 1-\exp\left(-N_e \left(m^\ell\right)^\gamma\right), \label{eq:rho}
\end{equation} calculated using \texttt{expm1()} to help avoid
underflow.

In keeping with the nomenclature introduced by
\citet{lawson2012inference}, we refer to \(h_i\) as the ``recipient
haplotype'' and the remaining haplotypes, \(h_{-i}\), as the ``donor
haplotypes'', in the context of the HMM where \(h_{i}\) is treated as
the emitted observation vector. This reflects the fact that each
recipient haplotype \(h_i\) is modelled as an imperfectly copied mosaic
of the other observed haplotypes under the LS model. Hence, the
posterior marginal probability at variant \(\ell\),
\(p^{\ell}_{ji} := \mathbb{P}\left(\left. X_i^\ell = j \right| h\right)\),
is the probability that donor \(j\) is copied by recipient \(i\) at
variant \(\ell\) given the haplotypes \(h\). Under the above definitions
of the prior copying probabilities \(\Pi\), the emission kernel
\eqref{eq:emission}, and the transition kernel \eqref{eq:transition},
the full \(N \times N\) matrix of copying probabilities at \(\ell\),
\(p^\ell\), can be obtained by running the standard forward and backward
recursions \citep{rabiner1989tutorial} for each column (ie for each
independent HMM).

From these posterior probabilities, we calculate a local \(N \times N\)
distance matrix, \(d^\ell\). Firstly, notice that theoretically
\(p_{ij}^\ell > 0\), but it can be that \(p_{ij}^\ell < \varepsilon\),
where \(\varepsilon\) is the double precision machine epsilon
(\(\approx 2.22\times10^{-16}\), \citet{C18}, pp.26). Effectively this
means \(d_{ij}^\ell\) is too large to reliably work with precisely, and
so for the purposes of distance calculations we treat \(\varepsilon\) as
the smallest observable posterior probability, yielding \begin{equation}
  d_{ji}^\ell = -\frac{\log\left(p_{ji}^\ell \vee \varepsilon \right) + \log\left(p_{ij}^\ell \vee \varepsilon  \right)}{2} \quad \forall\ j \neq i \label{eq:distmat}
\end{equation} where \(\vee\) is the maximum binary operator. By
convention \(d_{ii} = 0\) for all \(i\).

Please see \Cref{apx:nan} for some important discussion on parameter
values and numerical stability of the algorithm.

We proceed in next Section to reformulate the forward and backward
recursions so that we can more fully exploit modern high-performance CPU
instruction sets, while preserving numerical precision.

\hypertarget{sec:reformulation}{%
\subsection{Modification of the forward-backward
algorithm}\label{sec:reformulation}}

The \(N\) independent HMMs of the LS model then have forward and
backward probabilities, respectively: \begin{equation*}
  \tilde{\alpha}_{ji}^{\ell} = \mathbb{P}\left( X_i^\ell = j, h_i^{1:\ell} \right), \qquad 
  \tilde{\beta}_{ji}^{\ell} = \mathbb{P}\left(\left.  h_i^{\ell+1 : L}  \right| X_i^\ell = j \right), \qquad i \in \{1,\dots,N\} ,
\end{equation*} where \(h_{i}^{1:\ell}\) denotes haplotype \(i\) from
variant \(1\) to \(\ell\) inclusive.

Define, \begin{align}
  F_i^\ell &:= \sum_{j=1}^N \tilde{\alpha}_{ji}^{\ell} & F_i^{0} &:= 1 \label{eq:F} \\
  G_i^\ell &:= \sum_{j=1}^N \tilde{\beta}_{ji}^{\ell+1}\theta_{ji}^{\ell+1} \pi_{ji} & G_i^L &:= 1 \label{eq:G}
\end{align} Then the forward and backward recursions for the LS model
can be written in vector notation (subscript \(\cdot\) denoting the
vectorised index), \begin{align}
  \tilde{\alpha}_{\cdot i}^{\ell} &\gets \theta^{\ell}_{\cdot i} \left( \left(1-\rho^{\ell-1}\right) \tilde{\alpha}_{\cdot i}^{\ell-1} + \rho^{\ell-1} F_i^{\ell-1} \pi_{\cdot i} \right) & \text{for } \ell &\in \{2,\dots,L\}, \label{eq:raw_forward} \\
  \tilde{\beta}_{\cdot i}^{\ell} &\gets \left( 1- \rho^{\ell}\right) \tilde{\beta}_{\cdot i}^{\ell+1} \theta^{\ell+1}_{\cdot i} + \rho^{\ell} G_i^\ell & \text{for } \ell &\in \{1,\dots,L-1\} \label{eq:raw_backward}.
\end{align} with recursions initialised with
\(\alpha_{\cdot i}^{1} \gets \theta^{1}_{\cdot i} \pi_{\cdot i}\) and
\(\beta_{\cdot i}^{L} \gets 1\). Note that \Cref{eq:raw_forward}
corresponds to Equation A5 in \citet{Li2213}.

To partially mitigate the risk of underflow, the forward recursion can
be rearranged in terms of
\(\alpha_{\cdot i}^{\ell} := \frac{\tilde{\alpha}_{\cdot i}^{\ell}}{ F_i^{\ell-1}}\),
and the backward recursion in terms of
\(\beta_{\cdot i}^{\ell} := \frac{\tilde{\beta}_{\cdot i}^{\ell}}{G_i^{\ell}}\)
(see \Cref{apx:fwd,apx:bck} for details). Thus, in full for
\(\ell \in \{1,\dots,L\}\) we compute, \begin{align}
  \alpha_{\cdot i}^{1} &\gets \theta^{1}_{\cdot i} \pi_{\cdot i} &\textrm{for} \quad \ell = 1 \label{eq:fwd0} \\
  \alpha_{\cdot i}^{\ell} &\gets \theta^{\ell}_{\cdot i} \left( \left(1-\rho^{\ell-1}\right) \frac{\alpha_{\cdot i}^{\ell-1}}{\underset{j}{\sum} \alpha_{ji}^{\ell-1}} + \rho^{\ell-1} \pi_{\cdot i} \right) &\textrm{for} \quad \ell > 1 \label{eq:fwd1} \\ \intertext{and}
  \beta_{\cdot i}^{L} &\gets 1 &\textrm{for} \quad \ell = L \label{eq:bck0} \\
  \beta_{\cdot i}^{\ell} &\gets \left( 1- \rho^{\ell}\right) \frac{\beta_{\cdot i}^{\ell+1} \theta^{\ell+1}_{\cdot i}}{\underset{j}{\sum} \beta_{ji}^{\ell+1}\theta_{ji}^{\ell+1} \pi_{ji}} + \rho^\ell &\textrm{for} \quad \ell < L \label{eq:bck1}
\end{align}

Given \(\alpha_{\cdot i}^\ell\) and \(\beta_{\cdot i}^\ell\), the vector
of posterior probabilities for recipient \(i\), \(p_{\cdot i}\), can be
calculated directly by normalising, \begin{equation}
  p^\ell_{\cdot i} = \frac{\alpha^\ell_{\cdot i} \odot \beta^\ell_{\cdot i}}{\sum\limits_j \alpha^\ell_{ji} \odot \beta^\ell_{ji}} \label{eq:postprob}
\end{equation} where \(\odot\) denotes the Hadamard product. In the
event that \(\sum\limits_j \alpha^\ell_{ji} \odot \beta^\ell_{ji} = 0\),
the distance between the recipient haplotype \(i\) and all of the donor
haplotypes is beyond numerical precision, so as per the previous
sub-section we define \(p_{ji}^\ell = \varepsilon \ \forall\ j \ne i\).

Finally, the local distances follow by taking the negative log and
symmetrising. Note that if the distances are standardised for one of
these columns, to account for the fact that the standard deviation will
be 0, we set all of the standardised distances to 0.

Please see \Cref{apx:nan} for some discussion on parameter values and
exactly how \pkg{kalis} performs certain computations for numerical
stability of the algorithm.

\section[The kalis package]{The \pkg{kalis} package}\label{sec:thepkg}

\pkg{kalis} is an \proglang{R} package \citep{R}, with all time critical
code developed in \proglang{C} \citep{C18} with extensive use of
low-level SIMD instructions. These full technical details are presented
in \Cref{sec:core} for the interested reader, but from a user
perspective these high performance implementation details are hidden
behind a user-friendly API. In the remainder of this section we
introduce the package from a user perspective, from package installation
right through to decoding a single variant position.

\subsection[Installing kalis]{Installing
\pkg{kalis}}\label{sec:installing}

\pkg{kalis} will soon be available on CRAN, but users on MacOS and
Windows should be aware that out of necessity the CRAN binary will be
compiled for maximum compatibility, not maximum performance. We
\emph{strongly} recommend compiling the package from source code if you
plan to work with large haplotype sets (eg \(N > 1000\)).

\hypertarget{compiling-from-source-for-maximum-performance}{%
\subsubsection{Compiling from source for maximum
performance}\label{compiling-from-source-for-maximum-performance}}

To install directly from Github, compiling from source, is easiest by
using the \pkg{remotes} package \citep{remotes}. \emph{This is the
recommended installation method since bug fixes are pushed immediately
to Github.}

\begin{verbatim}
remotes::install_github("louisaslett/kalis",
  configure.vars = c(kalis = "PKG_CFLAGS='-march=native -mtune=native -O3'"))
\end{verbatim}

If the above installation command works correctly, the
\texttt{PKG\_CFLAGS} setting will be reported back to you. In most
cases, \pkg{kalis} will then be able to auto-detect the vector
instruction set of your CPU and this will be reported too. For example,
you would see the following near the start of the console output on a
modern Intel CPU which supports AVX2:

\begin{verbatim}
Using PKG_CFLAGS=-march=native -mtune=native -O3
Using PKG_LIBS=-lz 
AVX2 family of instruction set extentions will by used (auto-detected).
\end{verbatim}

If you have an error in passing the \texttt{PKG\_CFLAGS} setting, the
first line will instead read:

\begin{verbatim}
Using PKG_CFLAGS=
\end{verbatim}

If auto-detection of the instruction set has failed (or if the
\texttt{PKG\_CFLAGS} was not passed properly), then the third line will
instead read:

\begin{verbatim}
No special assembly instruction set extentions will by used (auto-detected).
\end{verbatim}

It is possible to override the auto-detection and manually direct
\pkg{kalis} which instruction set to use, please see
\Cref{apx:compileis} for details.

Finally, when you load the \pkg{kalis} package, you will also receive a
diagnostic message confirming the status of the compiled code. For
example, the aforementioned Intel CPU compilation provides the following
diagnostic on loading:

\begin{CodeChunk}
\begin{CodeInput}
R> library("kalis")
\end{CodeInput}
\begin{CodeOutput}

Running in 64-bit mode using x86-64 architecture.
Loops unrolled to depth 4.
Currently using AVX2, AVX, SSE4.1, SSE2, FMA and BMI2 CPU instruction set extensions.
\end{CodeOutput}
\end{CodeChunk}

Advanced users should note that \pkg{kalis} will respect \texttt{CFLAGS}
settings in \texttt{\textasciitilde{}/.R/Makevars}, so be alert to any
compiler flags set there that may conflict with the above installation
commands.

Advanced users may also be interested to benchmark performance under
different levels of loop unrolling: manually controlling this setting at
compile time is described in \Cref{apx:unroll}.

\hypertarget{package-overview}{%
\subsection{Package overview}\label{package-overview}}

In the \emph{v}1 release of \pkg{kalis} there are 18 functions which
enable computation of the model described in \Cref{sec:algorithm}. These
are briefly summarised in the following table, grouped by the task under
which the function falls.

\newpage

\begin{CodeChunk}
\begin{longtable}{ll}
\toprule
\textbf{Function} & \textbf{Purpose} \\ 
\midrule
\multicolumn{2}{l}{\emph{Load and Inspect Haplotypes}} \\ 
\midrule
\texttt{CacheHaplotypes} & Reads haplotype data into internal \pkg{kalis} format \\ 
\texttt{CacheSummary} & Prints current state of the haplotype cache \\ 
\texttt{ClearHaplotypeCache} & Frees internal cache memory \\ 
\texttt{N} and \texttt{L} & Retrieve number of phased haplotypes, $N$, and sites, $L$ \\ 
\texttt{QueryCache} & Retrieve haplotypes from the internal memory cache \\ 
\midrule
\multicolumn{2}{l}{\emph{Initialise HMM}} \\ 
\midrule
\texttt{MakeBackwardTable} & Constructs $N \times N$ backward matrix for $N$ HMMs \\ 
\texttt{MakeForwardTable} & Constructs $N \times N$ forward matrix for $N$ HMMs \\ 
\texttt{Parameters} & Define LS model parameters $\rho, \mu, \Pi$ and compute options \\ 
\midrule
\multicolumn{2}{l}{\emph{HMM Propagation}} \\ 
\midrule
\texttt{Backward} & Executes the backward recursion of \Cref{eq:bck0,eq:bck1} \\ 
\texttt{Forward} & Executes the forward recursion of \Cref{eq:fwd0,eq:fwd1} \\ 
\midrule
\multicolumn{2}{l}{\emph{Decode HMM}} \\ 
\midrule
\texttt{DistMat} & Computes distances per \Cref{eq:distmat} \\ 
\texttt{PostProbs} & Compute posterior marginal probabilites, \Cref{eq:postprob} \\ 
\midrule
\multicolumn{2}{l}{\emph{Utilities}} \\ 
\midrule
\texttt{CalcRho} & Compute recombination probabilities, $\rho$, \Cref{eq:rho} \\ 
\texttt{CopyTable} & Creates a fully cloned (deep) copy of table \\ 
\texttt{ReadHaplotypes} & Load haplotypes from HDF5 format into an R object \\ 
\texttt{ResetTable} & Efficiently wipe a forward/backward table for reuse \\ 
\texttt{WriteHaplotypes} & Save haplotypes from binary R matrix into HDF5 format \\ 
\bottomrule
\end{longtable}
\end{CodeChunk}

Full details of these functions can be found in the package
documentation, or at the package website,
\url{https://kalis.louisaslett.com/}.

In the remainder of this section we provide detailed comments on the
steps involved in using \pkg{kalis} to compute the posterior marginal
probabilities (\cref{eq:postprob}) or distances (\cref{eq:distmat}),
with numerous pertinent asides to aid using the package efficiently.
These steps are broken down as: (i) loading the haplotype data from
\proglang{R} or disk; (ii) setting the model parameters
\(\rho, \mu, \Pi\); (iii) initialising the \(N\) HMMs; (iv) running the
forward/backward algorithms; (v) computing the posterior marginal
probabilities and distances.

\hypertarget{sec:loadinghaps}{%
\subsection{Loading haplotype data}\label{sec:loadinghaps}}

In order to demonstrate how to use \pkg{kalis}, the package comes with a
toy data set of 300 simulated haplotypes, \texttt{SmallHaps}.

\begin{CodeChunk}
\begin{CodeInput}
R> library("kalis")
\end{CodeInput}
\begin{CodeOutput}

Running in 64-bit mode using x86-64 architecture.
Loops unrolled to depth 4.
Currently using AVX2, AVX, SSE4.1, SSE2, FMA and BMI2 CPU instruction set extensions.
\end{CodeOutput}
\begin{CodeInput}
R> data("SmallHaps")
\end{CodeInput}
\end{CodeChunk}

This simulated dataset is stored as an \(L = 400\) by \(N = 300\) matrix
with binary entries, as can be seen by inspecting with \texttt{str()},

\begin{CodeChunk}
\begin{CodeInput}
R> str(SmallHaps)
\end{CodeInput}
\begin{CodeOutput}
 int [1:400, 1:300] 0 0 0 0 0 0 0 1 0 0 ...
\end{CodeOutput}
\end{CodeChunk}

In order to run the LS model, haplotype data must first be loaded into
an internal optimised cache (see \Cref{sec:haplotypedata} for technical
details) using the \texttt{CacheHaplotypes()} function. This function
accepts an \proglang{R} matrix in this form, but also loading from
genetics file formats direct from disk (see next part).

\begin{CodeChunk}
\begin{CodeInput}
R> CacheHaplotypes(SmallHaps)
\end{CodeInput}
\end{CodeChunk}

The cache format is a raw binary representation that cannot be natively
viewed in \proglang{R}. Therefore, there is a utility function,
\texttt{QueryCache()}, to read all (or parts) of the cache into an
\proglang{R} matrix representation, enabling inspection to ensure the
data has loaded correctly.

For example, we can confirm that the internal cache contains
\texttt{SmallHaps} by retrieving summary information. With such a small
matrix we can also use \texttt{QueryCache()} to retrieve the whole
matrix from the internal cache (by passing no arguments), and
demonstrate that all entries match.

\begin{CodeChunk}
\begin{CodeInput}
R> CacheSummary()
\end{CodeInput}
\begin{CodeOutput}
Cache currently loaded with 300 haplotypes, each with 400 variants. 
  Memory consumed: 25.60 kB. 
\end{CodeOutput}
\begin{CodeInput}
R> all(QueryCache() == SmallHaps)
\end{CodeInput}
\begin{CodeOutput}
[1] TRUE
\end{CodeOutput}
\end{CodeChunk}

\texttt{QueryCache()} also supports retrieving just certain
variants/haplotypes by providing a numeric vector of required
variants/haplotypes respectively as the first two arguments (with
standard \proglang{R} \texttt{1} indexing), which is particularly useful
when using \pkg{kalis} for very large problem sizes where the whole
haplotype matrix is too big for \proglang{R}. Thus, the following code
compares just the first 10 haplotypes at sites 42 and 54.

\begin{CodeChunk}
\begin{CodeInput}
R> all(QueryCache(c(42, 54), 1:10) == SmallHaps[c(42, 54), 1:10])
\end{CodeInput}
\begin{CodeOutput}
[1] TRUE
\end{CodeOutput}
\end{CodeChunk}

Please note, it can be important to avoid mutations that appear on only
a single haplotype (singletons) for numerical stability, see
\Cref{apx:nan} for further discussion.

At this juncture we are, in principle, ready to move to discuss setting
parameters and then running the LS model. However, for practical
real-world problems it is rare that one would choose to load the genetic
data via an R matrix, so we first discuss supported file formats.

\hypertarget{recommendations-on-loading-haplotypes}{%
\subsubsection{Recommendations on loading
haplotypes}\label{recommendations-on-loading-haplotypes}}

With real-world large haplotype data sets it is preferable to avoid
having to load them into \proglang{R} at all, instead having \pkg{kalis}
read directly from disk into the internal optimised cache. Therefore,
\texttt{CacheHaplotypes()} also supports loading from two on-disk
formats directly, bypassing loading into \proglang{R} at all. In both
cases, a string containing the file name is passed, instead of an
\proglang{R} matrix.

Therefore, in all there are three supported methods for
\texttt{CacheHaplotypes()} to load haplotype data:

\begin{itemize}
\tightlist
\item
  As already covered, directly from an \proglang{R} matrix with
  \texttt{0} and \texttt{1} entries. The haplotypes should be stored in
  columns, with variants in rows. Once loaded in the cache, the matrix
  can be safely removed from the \proglang{R} environment since it is
  internally stored in \pkg{kalis} in a more efficient format.
\item
  From a \texttt{.hap.gz} file containing data in the HAP/LEGEND/SAMPLE
  format used by IMPUTE2 \citep{impute2} and SHAPEIT \citep{shapeit}.
\item
  From an HDF5 file \citep{hdf5}, with the format described in
  \Cref{apx:hdf5}.
\end{itemize}

If you have data in another format, \pkg{kalis} provides
\texttt{WriteHaplotypes()} and \texttt{ReadHaplotypes()} to work with
the HDF5 format. This means that one can loaded genetic data into an
\proglang{R} matrix by other means and then save into the native on-disk
format recommended for \pkg{kalis}. The advantage here is that once
written to disk, the \proglang{R} session can be restarted to eliminate
the inefficient matrix representation and the haplotypes loaded directly
from HDF5 to the optimised internal cache. See \Cref{apx:writehdf5} for
full details.

Finally, note that \pkg{kalis} will only cache and operate on one
haplotype data set at a time, since the software is designed for
operating at large scale. As a result, calling
\texttt{CacheHaplotypes()} a second time frees the allocated cache
memory and loads the new data set. Of course, there is no restriction in
loading \pkg{kalis} in multiple separate \proglang{R} processes on the
same machine, each caching different data sets.

\hypertarget{ls-model-parameters}{%
\subsection{LS model parameters}\label{ls-model-parameters}}

Recall from \Cref{sec:algorithm} that the LS model is parametrised by
\(\rho, \mu,\) and \(\Pi\). \pkg{kalis} bundles these parameters
together into an environment of class \texttt{kalisParameters}, which
can be created by using the \texttt{Parameters()} function. For full
technical details of why an environment is used and what this object is
like, see \Cref{sec:core}. The three key arguments to
\texttt{Parameters()} are:

\begin{description}
\item[\(\rho =\) \texttt{rho}]
This is a numeric vector parameter which must have length \(L-1\). Note
that element \texttt{i} of this vector should be the recombination
probability between variants \texttt{i} and \texttt{i+1}.

There is a utility function, \texttt{CalcRho()}, to assist with creating
these recombination probabilities from a recombination map, described
below.

By default, the recombination probabilities are set to zero everywhere.
\item[\(\mu =\) \texttt{mu}]
The mutation probabilities may be specified either as uniform across all
variants (by providing a single scalar value), or may vary at each
variant (by providing a vector of length \(L\)).

By default, mutation probabilities are set to \(10^{-8}\).
\item[\(\Pi =\) \texttt{Pi}]
The original \citet{Li2213} model assumed that each haplotype has an
equal prior probability of copying from any other. However, in the
spirit of ChromoPainter \citep{lawson2012inference} we allow a matrix of
prior copying probabilities.

The copying probabilities may be specified as a standard \proglang{R}
matrix of size \(N \times N\). The element at row \texttt{j}, column
\texttt{i} corresponds to the prior (background) probability that
haplotype \texttt{i} copies from haplotype \texttt{j}. Note that the
diagonal \emph{must} by definition be zero and columns \emph{must} sum
to one.

Alternatively, for uniform copying probabilities, this argument need not
be specified, resulting in copying probability \(\frac{1}{N-1}\)
everywhere by default.

\textbf{Note 1:} there is a computational cost associated with
non-uniform copying probabilities, so it is recommended to leave the
default of uniform probabilities when appropriate. This is achieved by
omitting this argument.

\textbf{Note 2:} do \emph{not} specify a uniform matrix when uniform
probabilities are intended, since this would end up incurring the
computational cost of non-uniform probabilities.
\end{description}

The \texttt{Parameters()} function accepts two further logical flag
arguments, \texttt{use.speidel} (default \texttt{FALSE}) and
\texttt{check.rho} (default \texttt{TRUE}). The former enables the
asymmetric mutation model in RELATE \citep{speidel}, while the latter
performs machine precision checks.

Thus, to create the default parameter set
(\(\rho^\ell = 0 \ \forall\,\ell\),
\(\mu^\ell = 10^{-8} \ \forall\,\ell\), and
\(\Pi_{ii} = 0 \ \forall\,i, \ \Pi_{ij} = (N-1)^{-1} \ \forall\,i \ne j\))
it suffices to simply call,

\begin{CodeChunk}
\begin{CodeInput}
R> pars <- Parameters()
R> pars
\end{CodeInput}
\begin{CodeOutput}
Parameters object with:
  rho   = (0, 0, 0, ..., 0, 0, 1)
  mu    = 1e-08
  Pi    = 0.00334448160535117 
\end{CodeOutput}
\end{CodeChunk}

where this continues the \texttt{SmallHaps} example started in
\Cref{sec:loadinghaps}, so \((300-1)^{-1} \approx 0.003344\). In
particular, note that \texttt{Parameters()} can only be invoked
\emph{after} the haplotype data is loaded by \texttt{CacheHaplotypes()}
since \texttt{Parameters()} checks that the parameter specifications are
consistent with the data.

Perhaps most commonly, one may want to use a particular value of
\(\rho\), based on a recombination map, according to \Cref{eq:rho}.
\texttt{CalcRho()} enables this, allowing the specification of a vector
of recombination distances in centimorgans (argument \texttt{cM}), as
well as the scalar multiple \(N_e\) and power \(\gamma\) (respectively
arguments \texttt{s} and \texttt{gamma}, both defaulting to \texttt{1}).
Continuing the \texttt{SmallHaps} example, the package ships with a
corresponding simulated recombination map in \texttt{SmallMap}. We can
go from recombination map to recombination distances using
\texttt{diff()} (and leave the default \(N_e\) and \(\gamma\)):

\begin{CodeChunk}
\begin{CodeInput}
R> data("SmallMap")
R> rho <- CalcRho(diff(SmallMap))
R> pars <- Parameters(rho)
R> pars
\end{CodeInput}
\begin{CodeOutput}
Parameters object with:
  rho   = (6.99999975500001e-08, 9.99999995000001e-09, 4.999999875e-08, ..., 3.99999992000053e-08, 1.39999990200002e-07, 1)
  mu    = 1e-08
  Pi    = 0.00334448160535117 
\end{CodeOutput}
\end{CodeChunk}

\hypertarget{setting-up-the-hmms}{%
\subsection{Setting up the HMMs}\label{setting-up-the-hmms}}

At this point of the analysis pipeline, the haplotypes are loaded into
the internal optimised cache, and the parameters for the LS model have
been defined. The final step before running the forward/backward
algorithm, is to setup the storage for the \(N\) independent HMM forward
and backward probabilities, \(\alpha^\ell_{\cdot i}\) and
\(\beta^\ell_{\cdot i}, i \in \{1, \dots, N\}\).

\pkg{kalis} uses \(N \times N\) matrices wrapped in list objects which
are of class \texttt{kalisForwardTable} and \texttt{kalisBackwardTable}
respectively to store the forward/backward probabilities at a variant
\(\ell\). These objects additionally contain which site, \(\ell\), the
matrix represents and for performance reasons also retains the vector of
\(N\) scaling constants associated with the \(N\) HMMs (corresponding to
\Cref{eq:F} for the forward and \Cref{eq:G} for the backward). These
objects can be created with \texttt{MakeForwardTable()} and
\texttt{MakeBackwardTable()}, and at a minimum the parameters of the LS
model to be used must be supplied.

Hereinafter, we refer to the collective contents of these as the
`forward table', `backward table', or simply `table'.

Continuing the \texttt{SmallHaps} example,

\begin{CodeChunk}
\begin{CodeInput}
R> fwd <- MakeForwardTable(pars)
R> fwd
\end{CodeInput}
\begin{CodeOutput}
Full Forward Table object for 300 haplotypes. 
  Newly created table, currently uninitialised to any variant (ready for Forward function next).
  Memory consumed: 723.98 kB 
\end{CodeOutput}
\begin{CodeInput}
R> bck <- MakeBackwardTable(pars)
R> bck
\end{CodeInput}
\begin{CodeOutput}
Full Backward Table object for 300 haplotypes, in rescaled probability space. 
  Newly created table, currently uninitialised to any variant (ready for Backward function next).
  Memory consumed: 724.13 kB 
\end{CodeOutput}
\end{CodeChunk}

These forward and backward tables of HMMs are now ready to be
`propagated' along the genome.

\hypertarget{sec:runningls}{%
\subsection{Running the LS model}\label{sec:runningls}}

Everything is in place now to run the LS model. The forward
\cref{eq:fwd0,eq:fwd1}, and backward \cref{eq:bck0,eq:bck1}, can now be
executed by using the \texttt{Forward()} and \texttt{Backward()}
functions respectively. These must both, as a minimum, be supplied with
a corresponding forward/backward table and with the parameters of the
model (in each of the first two arguments, \texttt{fwd}/\texttt{bck}
respectively and \texttt{pars}). By default this will result in a single
variant move either forwards or backwards. Moves directly to a
designated variant can be achieved by specifying the third argument
(\texttt{t}).

\textbf{Note:} that \pkg{kalis} seeks to minimize memory creation and
copying for increased performance, and so the tables supplied to
\texttt{Forward()} and \texttt{Backward()} \emph{are modified in place}.

Continuing the \texttt{SmallHaps} example, we can see this in action,
whereby the \texttt{fwd} table created in the previous subsection is
propagated without assignment, first initialising to \(\ell=1\), then
moving to \(\ell=2\):

\begin{CodeChunk}
\begin{CodeInput}
R> Forward(fwd, pars)
R> fwd
\end{CodeInput}
\begin{CodeOutput}
Full Forward Table object for 300 haplotypes. 
  Current variant = 1 
  Memory consumed: 723.98 kB 
\end{CodeOutput}
\begin{CodeInput}
R> Forward(fwd, pars)
R> fwd
\end{CodeInput}
\begin{CodeOutput}
Full Forward Table object for 300 haplotypes. 
  Current variant = 2 
  Memory consumed: 723.98 kB 
\end{CodeOutput}
\end{CodeChunk}

Likewise we can propagate according to the backward equations:

\begin{CodeChunk}
\begin{CodeInput}
R> Backward(bck, pars)
R> bck
\end{CodeInput}
\begin{CodeOutput}
Full Backward Table object for 300 haplotypes, in rescaled probability space. 
  Current variant = 400 
  Memory consumed: 724.13 kB 
\end{CodeOutput}
\begin{CodeInput}
R> Backward(bck, pars)
R> bck
\end{CodeInput}
\begin{CodeOutput}
Full Backward Table object for 300 haplotypes, in rescaled probability space. 
  Current variant = 399 
  Memory consumed: 724.13 kB 
\end{CodeOutput}
\end{CodeChunk}

\hypertarget{in-place-modification-of-tables}{%
\subsubsection{In-place modification of
tables}\label{in-place-modification-of-tables}}

A few additional comments on the in-place modification behaviour are in
order since this is contrary to the idiomatic \proglang{R} style, where
arguments are usually unaffected by function calls. The convention in
\proglang{R} would typically be that if an argument is to be updated, it
is copied inside the function and the modified copy returned: this would
incur an unacceptably high performance penalty in many intended use
cases for \pkg{kalis}, hence in place modification of the tables.
Indeed, for particularly large \(N\) problems, it may only be possible
to hold a few \(N \times N\) tables in memory at once so duplication is
impossible.

This has important knock-on ramifications if the user wishes to create a
duplicate of the table, due to the copy-on-write semantics of
\proglang{R}. As stated in the R Internals documentation
\citep[\S1.1.2]{Rinternals}:

\begin{quote}
``R has a `call by value' illusion, so an assignment like
\texttt{b\ \textless{}-\ a} appears to make a copy of \texttt{a} and
refer to it as \texttt{b}. However, if neither \texttt{a} nor \texttt{b}
are subsequently altered there is no need to copy. {[}\ldots{]} When an
object is about to be altered {[}\ldots{]} the object must be duplicated
before being changed.''
\end{quote}

The low level in-place modification by \pkg{kalis} does not alert the
\proglang{R} memory manager, meaning that a user who executes the
following will have unexpected results:

\begin{CodeChunk}
\begin{CodeInput}
R> fwd2 <- fwd
R> Forward(fwd2, pars)
\end{CodeInput}
\end{CodeChunk}

One might expect \texttt{fwd2} to now be 1 variant further advanced than
\texttt{fwd}, but in fact both are references to the \emph{same} object,
so both appear to have moved a variant. For this reason, \pkg{kalis}
provides the \texttt{CopyTable()} function which is the only supported
method of copying the content of a table. The order of arguments in
\texttt{CopyTable()} is designed to mimic assignment, with the
destination on the left.

Again, in the interest of minimising functions which cause memory
allocation, \texttt{CopyTable()} copies between \emph{existing} tables.
Therefore to achieve the effect intended in the previous code snippet,
the correct approach is to create a destination table and then copy:

\begin{CodeChunk}
\begin{CodeInput}
R> fwd2 <- MakeForwardTable(pars)
R> CopyTable(fwd2, fwd)
R> Forward(fwd2, pars)
\end{CodeInput}
\end{CodeChunk}

After these three lines are executed, \texttt{fwd2} will be a forward
table propagated one variant further than \texttt{fwd}.

\hypertarget{decoding-a-single-variant}{%
\subsection{Decoding a single variant}\label{decoding-a-single-variant}}

In practice, the objective of an analysis using the LS model is to
propagate both a forward and backward table of \(N\) HMMs to a common
variant \(\ell\), then compute the posterior marginal probabilities
(\cref{eq:postprob}) or distances (\cref{eq:distmat}).

Continuing our \texttt{SmallHaps} example, let us compute both
quantities at \(\ell=250\). This is now straight-forward, as we first
propagate directly to the destination \(\ell\),

\begin{CodeChunk}
\begin{CodeInput}
R> Forward(fwd, pars, 250)
R> Backward(bck, pars, 250)
R> fwd
\end{CodeInput}
\begin{CodeOutput}
Full Forward Table object for 300 haplotypes. 
  Current variant = 250 
  Memory consumed: 723.98 kB 
\end{CodeOutput}
\begin{CodeInput}
R> bck
\end{CodeInput}
\begin{CodeOutput}
Full Backward Table object for 300 haplotypes, in rescaled probability space. 
  Current variant = 250 
  Memory consumed: 724.13 kB 
\end{CodeOutput}
\end{CodeChunk}

Now that \texttt{fwd} and \texttt{bck} are at the same variant, they can
be combined to obtain \(p^\ell\) or \(d^\ell\), and the distance matrix
plotted:

\begin{CodeChunk}
\begin{CodeInput}
R> p <- PostProbs(fwd, bck)
R> d <- DistMat(fwd, bck)
R> plot(d)
\end{CodeInput}


\begin{center}\includegraphics{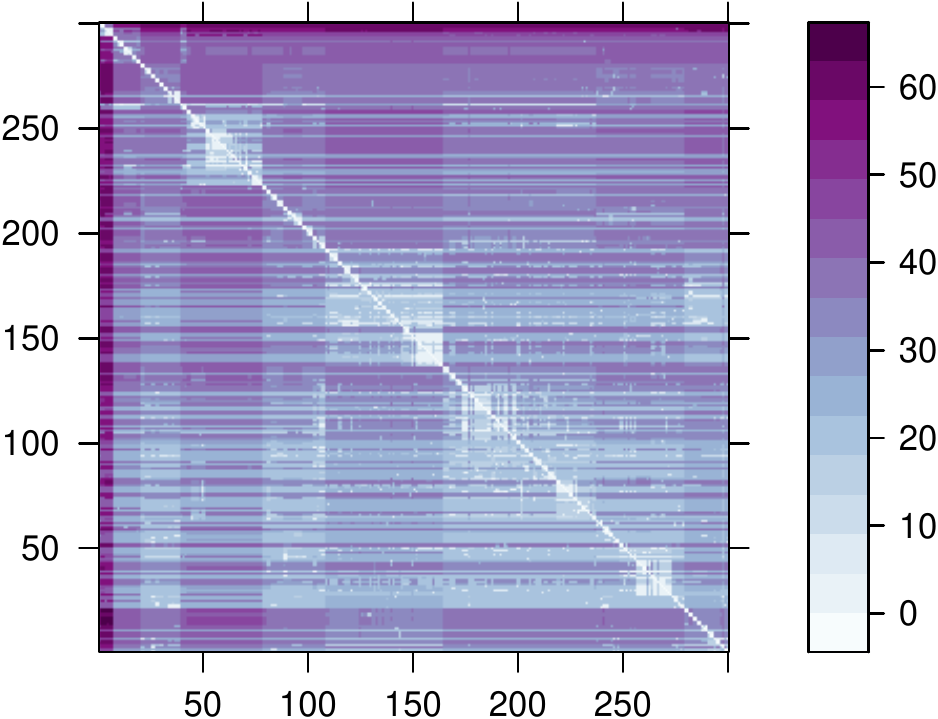} \end{center}

\end{CodeChunk}

\hypertarget{summary}{%
\subsection{Summary}\label{summary}}

This section has carefully walked through the pipeline involved in
running \pkg{kalis} for the LS model. The following code is the full
pipeline brought together for readability, without asides:

\begin{CodeChunk}
\begin{CodeInput}
R> library("kalis")
R> data("SmallHaps")
R> data("SmallMap")
R> 
R> CacheHaplotypes(SmallHaps)
R> 
R> rho <- CalcRho(diff(SmallMap))
R> pars <- Parameters(rho)
R> 
R> fwd <- MakeForwardTable(pars)
R> bck <- MakeBackwardTable(pars)
R> 
R> Forward(fwd, pars, 250)
R> Backward(bck, pars, 250)
R> 
R> p <- PostProbs(fwd, bck)
R> d <- DistMat(fwd, bck)
\end{CodeInput}
\end{CodeChunk}

\newpage

\hypertarget{sec:advtopics}{%
\subsection{Advanced topics}\label{sec:advtopics}}

There are a couple of `advanced' topics worth mentioning which are
available in the \pkg{kalis} API.

\hypertarget{handling-massive-haplotype-datasets}{%
\subsubsection{Handling massive haplotype
datasets}\label{handling-massive-haplotype-datasets}}

Firstly, in order to support massive haplotype sizes, it is possible
when making a table to create only a window of recipients (ie for some
subset of \(i \in \{1, \dots, N\}\)). The HMMs are all independent, so
this facilitates propagating batches of columns of the full
\(N \times N\) matrix of independent HMMs on different machines without
requiring network communication. By default, \texttt{MakeForwardTable()}
and \texttt{MakeBackwardTable()} will include all recipients, but a
range can be specified with the arguments \texttt{from\_recipient} and
\texttt{to\_recipient}.

For example, in a problem where \(N = 100,000\), a single table would
require just over 80GB of memory. If one were working with a cluster
where machines only have 32GB of RAM each, then one might choose to work
with batches of \(12,500\) recipients (ie columns) and propagate them
independently on each machine, since a table then requires just over
10GB. For the forward table, this would be created by,

\begin{CodeChunk}
\begin{CodeInput}
R> # Machine 1
R> fwd <- MakeForwardTable(pars, 1, 12500)
R> # Machine 2
R> fwd <- MakeForwardTable(pars, 12501, 25000)
R> # ...
R> # Machine 8
R> fwd <- MakeForwardTable(pars, 87501, 100000)
\end{CodeInput}
\end{CodeChunk}

Likewise for the backward tables. All other operations are unaffected,
from caching through to forward/backward propagation.

Notice that the posterior probabilities and distance matrices are easily
computed in a distributed manner, involving only a Hadamard product and
single reduction operation, so that the final results can be computed
entirely distributed without having to actually form a
\(100,000 \times 100,000\) matrix in memory.

\hypertarget{fine-control-of-parallelism}{%
\subsubsection{Fine control of
parallelism}\label{fine-control-of-parallelism}}

Although the SIMD instruction set is fixed at compile time, as discussed
in \Cref{sec:installing}, the user can of course control the degree of
threading at run-time.

Both \texttt{Forward()} and \texttt{Backward()} functions accept an
\texttt{nthreads} argument. By default this uses the core \proglang{R}
\pkg{parallel} package to automatically detect the number of available
cores and uses all of them. However, as is common in multi-threaded
packages, the user can also specify a scalar value here to use a
different degree of parallelism.

There is a further option of interest to advanced users. If an integer
vector is supplied, instead of a single scalar value, then \pkg{kalis}
will attempt to pin threads to the corresponding core number, if the
platform you are using supports thread affinity. This can be
particularly useful on massive non-uniform memory access (NUMA) systems,
where different cores have different speed of access to different parts
of memory (cores are split over `nodes' and while they can access all
memory, it will be faster to access memory on the same node). On a Linux
system, any memory allocation will by preference try to be allocated on
the same node as the core. Thus, an advanced user can opt to launch as
many \proglang{R} processes as there are NUMA nodes and use
\texttt{taskset} on Linux to pin one process to one core on each NUMA
node. Then, by availing of the table recipient windowing described
above, these tables memory allocations will be occur on the node local
to each core. Finally, when calling
\texttt{Forward()}/\texttt{Backward()}, a vector of only the cores on
the corresponding NUMA node can be provided to \texttt{nthreads}, so
that essentially all cross-NUMA node memory accesses are eliminated for
maximum performance.

\hypertarget{sec:core}{%
\section{Core implementation}\label{sec:core}}

The \proglang{R} interface described hereinbefore is a thin wrapper
layer around a high-performance implementation of the core algorithm
which is written in standards compliant C18 \citep{C18}. Most data
structures are represented with native \proglang{R} types enabling user
inspection and manipulation, except for the haplotype sequences
themselves.

Computationally, the innermost forward and backward recursions are
implemented using compiler intrinsics to exploit a variety of modern CPU
instruction sets, including Streaming SIMD Extensions (SSE2 and SSE4.1),
Advanced Vector Extensions (AVX, AVX2, AVX-512 and FMA) and Bit
Manipulation Instructions (BMI2) on Intel platforms; as well as NEON on
ARM platforms. AVX2 is supported in Intel CPUs of the Haswell generation
(released Q2 of 2013) or later, AVX-512 tends to be available only in
recent Intel server and workstation grade CPUs, and NEON is available
for ARM Cortex-A and Cortex-R series CPUs, as well as Apple M1/M2 and
Amazon Web Services Graviton processors. Although this covers most CPUs
likely to be in use today, we none-the-less provide reference
implementations in pure standards compliant \proglang{C} which will
operate on any CPU architecture with a C18 compliant compiler. During
package compilation, the correct code-paths are compiled based on
detection of the presence or absence of the required instruction sets,
or at the direction of the user via compiler flags. See
\Cref{apx:compileis}.

It may be worth noting at this juncture that it was an explicit design
choice to target CPUs and not GPU or tensor cards initially. This is
because most University high performance computing clusters have
plentiful CPU resources, often with untapped power in advanced SIMD
instructions sets. We believe that the problem size that can be
realistically tackled in many genetics studies can be massively
increased \emph{without} needing to resort to add-on cards, though to
scale beyond even this we may explore heterogeneous computing
architectures in future \pkg{kalis} research.

In this section, we now describe the inner workings and design
principles of the package, first covering in detail the data structures
(both user facing and internal), followed by the computational
implementation.

\hypertarget{data-structures}{%
\subsection{Data structures}\label{data-structures}}

There are three user accessible data structures utilised in the package
and a low level binary haplotype representation which is not directly
user accessible. The principle data structures of interest to users are
forward and backward table objects, represented as native \proglang{R}
lists with respective S3 class names \texttt{kalisForwardTable} and
\texttt{kalisBackwardTable} (detailed in Table \ref{tab:fwdbck}), which
are created with package functions \texttt{MakeForwardTable()} and
\texttt{MakeBackwardTable()} respectively. The third user accessible
data structure holds the LS model parameters, represented as a native
\proglang{R} environment with S3 class name \texttt{kalisParameters},
which can be created with the package function \texttt{Parameters()}.

\hypertarget{sec:haplotypedata}{%
\subsubsection{Haplotype data}\label{sec:haplotypedata}}

The haplotypes are stored in an optimised binary representation which is
only natively accessible from within C. Note that here ``optimised'' is
not a reference space-optimisation: it would be possible to represent
the haplotypes in an even more compressed manner, but we aim for
streaming compute speed optimisation instead.

The haplotypes are loaded from disk and transformed to an in memory
cache in this representation via \texttt{CacheHaplotypes()}, but this
function does not return any handle to the loaded data. Thus the package
provides the accessor function \texttt{QueryCache()}, which copies
genome segments from the binary representation into native \proglang{R}
integer vectors for user inspection.

\begin{figure}
\centerline{\includegraphics[width=0.7\textwidth]{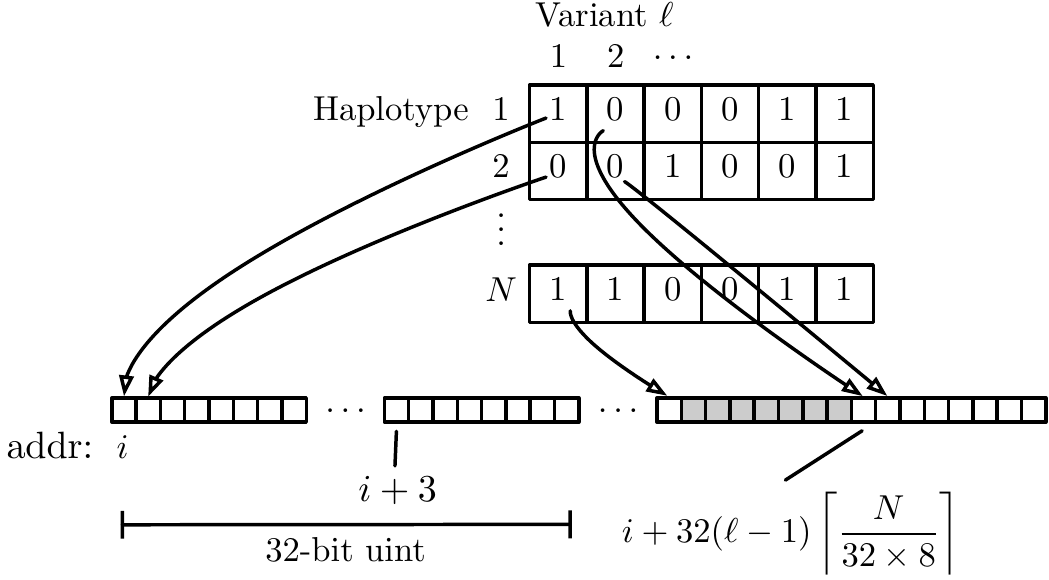}}
\caption{Efficient binary representation of interleaved haplotypes in memory, with 32-byte boundary alignment for each variant start for SSE/AVX instructions (here $i \mod 32 = 0$). The grey boxes indicate essentially 'wasted' bits which are ignored to ensure alignment for the start of the next variant.}
\label{fig:hapsmem}
\end{figure}

When \texttt{CacheHaplotypes()} loads haplotypes into the cache, they
are interleaved into a flat memory space which is organised as
variant-major. That is, variant 1 of each haplotype is loaded, converted
to a binary 0/1 and then 32 consecutive haplotypes are packed into an
unsigned integer. Moreover, the initial flat memory allocation is
aligned on a 32-byte boundary to satisfy memory alignment requirements
for some CPU vector
instructions\footnote{Certain modern CPUs do not require specific alignment to be able to load memory to SSE/AVX registers, but for maximum compatibility we honor the alignment anyway.},
and after all haplotypes at a given variant are packed into consecutive
unsigned integers the pointer is wound forward to the next 32-byte
boundary to ensure the next variant starts on an SSE/AVX vector
compatible memory boundary. This is depicted in Figure
\ref{fig:hapsmem}.

Firstly, note that this orientation is natural, since the forward and
backward recursions operate variant by variant, meaning variant-major
storage ensures sequential memory locations are being fetched during a
recursion. Indeed, with the cache line size of 64-bytes (starting Intel
Pentium IV), we essentially trigger the loading of \(64 \times 8 = 512\)
neighbouring variants upon accessing the first variant in a recursion.
This effect is even more pronounced on Apple M1/M2 whose cache line size
is 128-bytes, resulting in 1024 variants being pre-fetched upon access
to the first variant in a recursion.

Secondly, a possible drawback is that we must extract the individual bit
into a double floating point representation in order to compute with it
in the recursion. However, efficient CPU instructions can help here too:
take for example the following strategy \pkg{kalis} uses on an AVX2
capable CPU. Using the \texttt{PDEP} instruction in BMI2, we can
efficiently deposit a bit into every ninth bit of an \texttt{int} (so
there are now 4 8-bit integers taking on the value of the haplotype at
this variant packed in an \texttt{int}). Then, using SSE2, SSE4.1 and
AVX instructions one can inflate through representations from 4 8-bit
integers packed in an \texttt{int} up to 4 64-bit doubles packed in a
256-bit AVX register. As such, we are then ready to operate with this
variant in parallel using AVX instructions.

During development, testing indicated the memory bandwidth and cache
efficiency savings of the packed binary representation provided
speed-ups thanks to these instructions efficiently enabling unpacking
and spreading a haplotype variant bit for parallel use. Furthermore,
such a compact representation means that more of L1/L2 cache and memory
bus bandwidth is left available for forward and backward tables, which
are the largest objects we work with in this problem.

\hypertarget{parameters}{%
\subsubsection{Parameters}\label{parameters}}

The \texttt{kalisParameters} object uses an environment rather than list
for parameters for two reasons: (i) the parameter environment and its
bindings are locked which prevents changes in parameter values between
forward or backward table propagation steps, since parameters must be
fixed for all steps of a given forward or backward computation; and (ii)
an environment explicitly ensures the (often large) parameter vectors
are not copied when associated with potentially many different tables,
but will always be purely referenced.

The environment contains only two members: another environment with the
actual parameter values (which is locked with
\texttt{lockEnvironment()}); and a SHA-256 hash of those parameter
values (details in Table \ref{tab:pars}). The purpose of the hash is to
be able to efficiently determine whether the correct parameter set for a
given forward or backward table has been passed when computing forward
or backward recursions from an already initialised table (since it would
be incorrect to propagate forward or backward using different parameter
sets in different parts of the genome).

\begin{table}[tbp]
\centering
\begin{tabular}{l|ll}
  \hline
  \code{kalisParameters} object & Data type & \\ \hline\hline
  \code{pars} & Locked \proglang{R} environment, containing: \\
  & \code{rho} & vector length $L$ \\
  & \code{mu} & vector length $L$, or scalar \\
  & \code{Pi} & $N \times N$ matrix, or scalar \\
  \code{sha256} & character & \\ \hline
\end{tabular}
\caption{The content of the data structure representing parameter objects.}
\label{tab:pars}
\end{table}

\newpage

\hypertarget{forwardbackward-tables}{%
\subsubsection{Forward/backward tables}\label{forwardbackward-tables}}

\begin{table}[tbp]
\centering
\begin{tabular}{ll|ll|l}
  \hline
  \multicolumn{2}{l|}{\code{kalisForwardTable} object} &   \multicolumn{2}{l|}{\code{kalisBackwardTable} object} & Data type \\ \hline\hline
  \code{alpha} & $= \alpha^\ell_{\cdot\cdot}$ & \code{beta} & $= \beta^\ell_{\cdot\cdot}$  & $N   \times N$ matrix \\
  \code{alpha.f} & $= F^\ell$ & \code{beta.g} & $= G^\ell$ & vector length $N$ \\
  \code{l} & $= \ell$ & \code{l} & $= \ell$ & integer scalar \\
  \code{from_recipient} & & \code{from_recipient} & & integer scalar \\
  \code{to_recipient} & & \code{to_recipient} & & integer scalar \\
  \code{pars.sha256} & & \code{pars.sha256} & & character \\
  & & \code{beta.theta} & & logical scalar \\ \hline
\end{tabular}
\caption{The content of the core data structures representing forward and backward table objects, together with their correspondence to mathematical quantities defined in Section \ref{sec:algorithm}.}
\label{tab:fwdbck}
\end{table}

The forward and backward table objects contain not only the (upto) \(N\)
independent forward/backward tables at variant \(\ell\), but also
supporting supporting information. This includes the variant the table
is currently at, the scaling constants \(F^\ell\) (forward, \Cref{eq:F})
or \(G^\ell\) (backward, \Cref{eq:G}), the range of recipient haplotypes
represented (that is, the recipient HMMs to which the column
corresponds) as discussed in \Cref{sec:advtopics}, and a hash of the
parameter values used in propagating this table.

In total, a full-size forward table for example requires
\(8N^2+8N+1576\) bytes of
memory\footnote{Measured under \proglang{R} 4.2.2} for storage and the
small overhead of \proglang{R} object management. Since this grows
quadratically in the number of haplotypes, most functions in the package
operate on forward and backward table objects in-place, rather than via
the idiomatic copy-on-write mechanism of standard \proglang{R}, as
discussed in \Cref{sec:runningls}. The most important consequence of
this for users is that standard assignment of a table object to another
variable name only creates a reference and so an explicit copy must be
made by using the \texttt{CopyTable()} utility function provided (see
examples in that earlier Section).

\hypertarget{core-simd-code}{%
\subsection{Core SIMD code}\label{core-simd-code}}

The two most important core algorithms which are accelerated with SIMD
vector instructions are the forward and backward recursions. This code
is fully implemented in \proglang{C}, with tailored modifications
accounting for all combinations of scalar/vector \(\mu\), scalar/matrix
\(\Pi\), and \citet{speidel} or not (ie 8 combinations), to ensure that
minimal memory accesses are performed where possible. So, for example,
scalar \(\mu\) and scalar \(\Pi\) parameters will be faster than any
other combination since these values are likely to be held in registers
(or at least L1 cache) for the duration of the recursion.

Additionally, in all places where we identify SIMD instructions may be
used, a macro is deployed, with a header file providing all mappings
from these macros to a specific SIMD instruction for all supported
instruction sets. Taking arguably the simplest non-trivial example, all
\texttt{src/ExactForward*.c} and \texttt{src/ExactBackward*.c} files
make us of the custom macro \texttt{KALIS\_MUL\_DOUBLE(X,\ Y)} when they
need to multiply \texttt{KALIS\_DOUBLEVEC\_SIZE} double precision
floating point values together. The file, \texttt{src/StencilVec.h} then
provides definitions for these macros under each instruction set
\pkg{kalis} supports (via assembly intrinsics), together with a pure
\proglang{C} alternative. For this example, we have (with \texttt{...}
indicating other macro definitions):

\begin{verbatim}
// Extract from src/StencilVec.h
#if defined(KALIS_ISA_AVX512)
#define KALIS_DOUBLEVEC_SIZE 8
#define KALIS_MUL_DOUBLE(X, Y) _mm512_mul_pd(X, Y)
...
#elif defined(KALIS_ISA_AVX2)
#define KALIS_DOUBLEVEC_SIZE 4
#define KALIS_MUL_DOUBLE(X, Y) _mm256_mul_pd(X, Y)
...
#elif defined(KALIS_ISA_NEON)
#define KALIS_DOUBLEVEC_SIZE 2
#define KALIS_MUL_DOUBLE(X, Y) vmulq_f64(X, Y)
...
#elif defined(KALIS_ISA_NOASM)
#define KALIS_DOUBLEVEC_SIZE 1
#define KALIS_MUL_DOUBLE(X, Y) (X) * (Y)
...
#endif
\end{verbatim}

The inner-most loop in these core files then includes a programmatically
generated unroll to the depth specified during compilation. All this is
wrapped in code which dispatches using \texttt{pthreads} to multiple
threads, with automatic detection of the ability to pin to specific
cores if that option is passed (see \Cref{sec:advtopics}). In
particular, each thread operates on a subset of columns of the forward
and backward tables, ensuring spatial locality for memory accesses.
Furthermore, when propagating by more than a single variant position,
each column (ie each independent HMM) is propagated all the way to the
target variant before proceeding to the next column, ensuing temporal
locality of memory accesses.

\hypertarget{sec:perf}{%
\section{Performance}\label{sec:perf}}

We provide a brief overview of some example performance figures, though
due to the highly tuned nature of \pkg{kalis}, the exact performance you
can expect will be heavily dependent on your exact computer architecture
and resources.

\begin{figure}
  \centering
\hfill\includegraphics[width=0.45\textwidth]{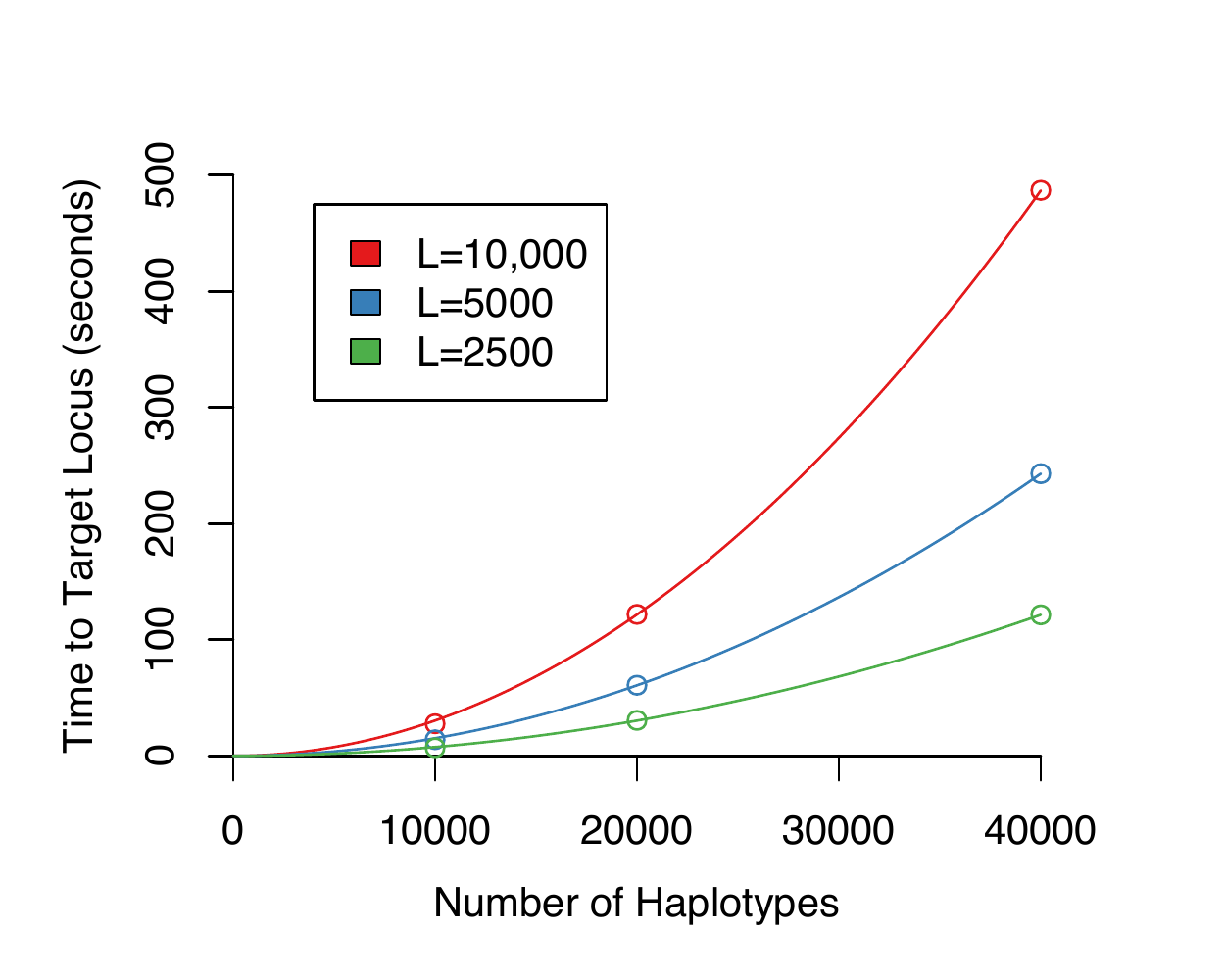}\hfill\includegraphics[width=0.45\textwidth]{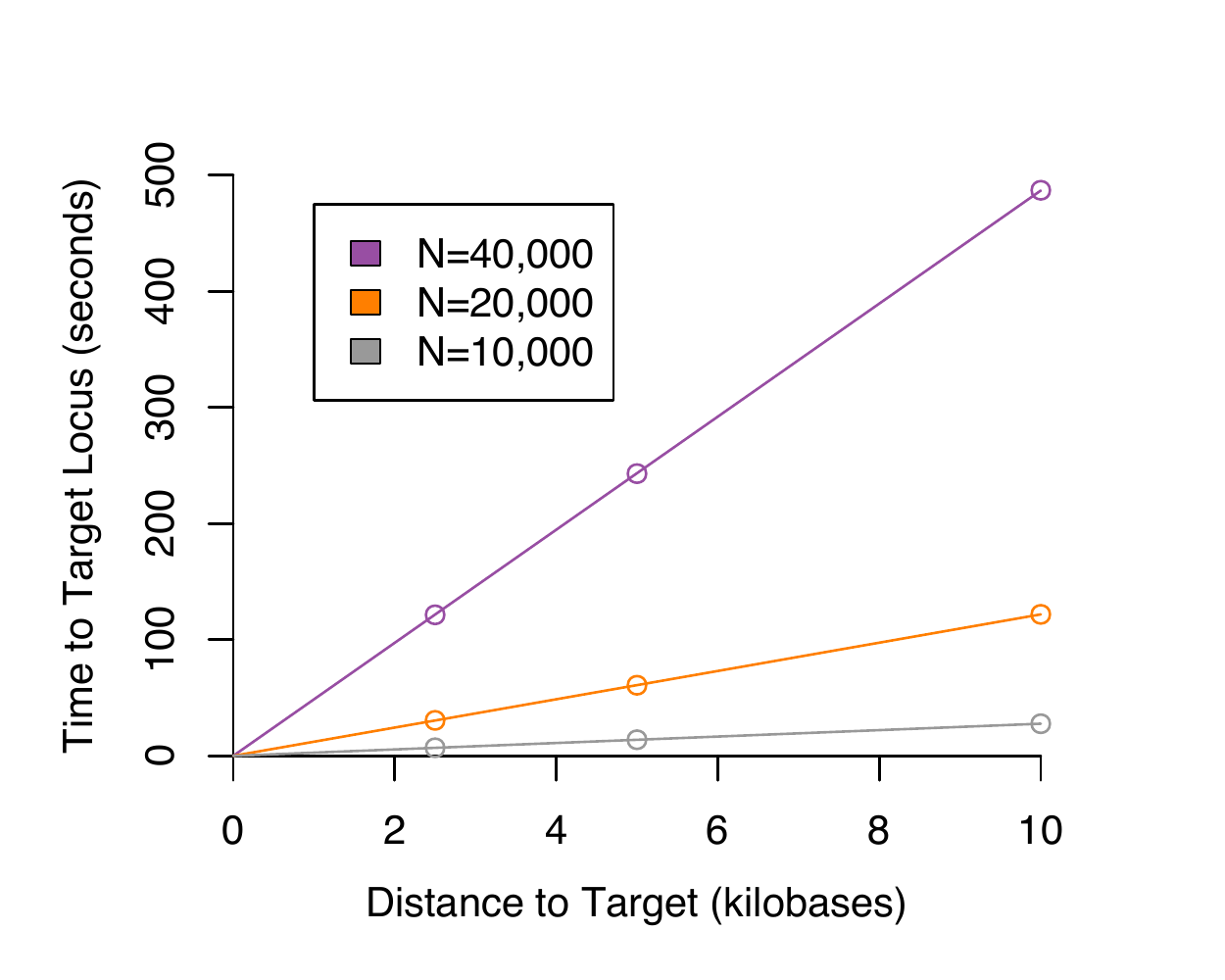}\hfill
    \caption{\pkg{kalis} shows the expected order $N^2$ and order $L$ scaling of the LS model.  Computed on an Amazon Web Services \texttt{c4.8xlarge} instance (36 vCPUs, 60 GB of RAM).}
  \label{fig:perfscaling}
\end{figure}

First, it is important to note we do \emph{not} claim to have altered
the scaling properties of the LS model, only that we provide an
implementation which is highly optimised within the scaling constraints
inherent to the model. As such, \Cref{fig:perfscaling} demonstrates that
\pkg{kalis} indeed inherits the \(\mathcal{O}(N^2)\) and
\(\mathcal{O}(L)\) properties of the original LS model.

We turn now to the benefits \pkg{kalis} does provide.

\begin{figure}
  \centering
  \includegraphics[width=0.95\textwidth]{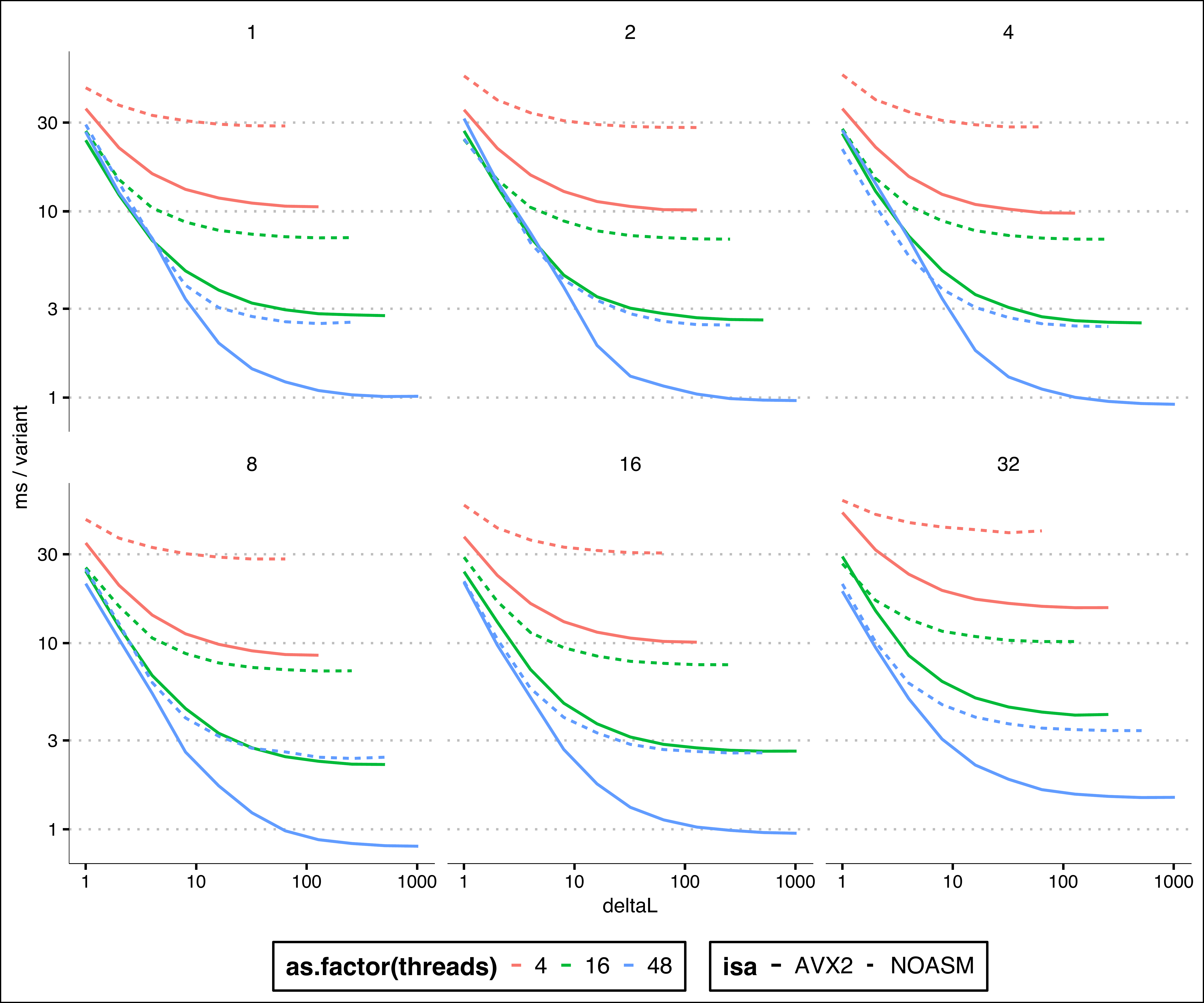}
    \caption{Log-log plot of milliseconds per variant performance ($y$-axis) of the forward algorithm on 10,000 haplotypes, against the number of variants propagated ($x$-axis). Each panel is a different loop unrolling depth (panel title gives loop unrolling level).  Line colour denotes number of CPU threads, whilst a dashed line indicates vanilla \proglang{C} and a solid line indicates hand-coded AVX2 instructions.  In total, using AVX2, 48 threads, and loop unrolls to depth 8, it takes less than 10 seconds to propagate a $10000 \times 10000$ forward table over 10,000 variants.}
  \label{fig:perfdeltaL1}
\end{figure}

Firstly, for some of the reasons highlighted in the previous Section,
\pkg{kalis} exhibits accelerated performance when propagating the
forward/backward recursions over more extended stretches of the genome.
This is because every effort has been made to be cache efficient, so
that when more than a single variant step is taken, the strong cache
locality design ensures that we are not memory bandwidth limited. This
effect can be seen quite dramatically in \Cref{fig:perfdeltaL1} by the
rapid decrease in compute time per variant as longer stretches are
propagated.

Secondly, the hard-coded loop unrolling functionality which can be
controlled at compile time by the user can be seen to be beneficial in
\Cref{fig:perfdeltaL1}. Clearly excessive loop unrolling is harmful,
with depth 32 unrolls actually being substantially slower than no
unrolling. However, unrolling to depth 8 does give a clear improvement.
The best choice of unrolls will be both problem and architecture
dependent, so we recommend testing different unroll levels on the target
problem before performing long compute runs.

\Cref{fig:perfdeltaL1} also illustrates that the hand-designed use of
low-level vector SIMD instructions is not superfluous, with substantial
speed-up afforded by their use (the difference between dashed and solid
lines of the same colour).

\begin{figure}
  \centering
  \includegraphics[width=0.95\textwidth]{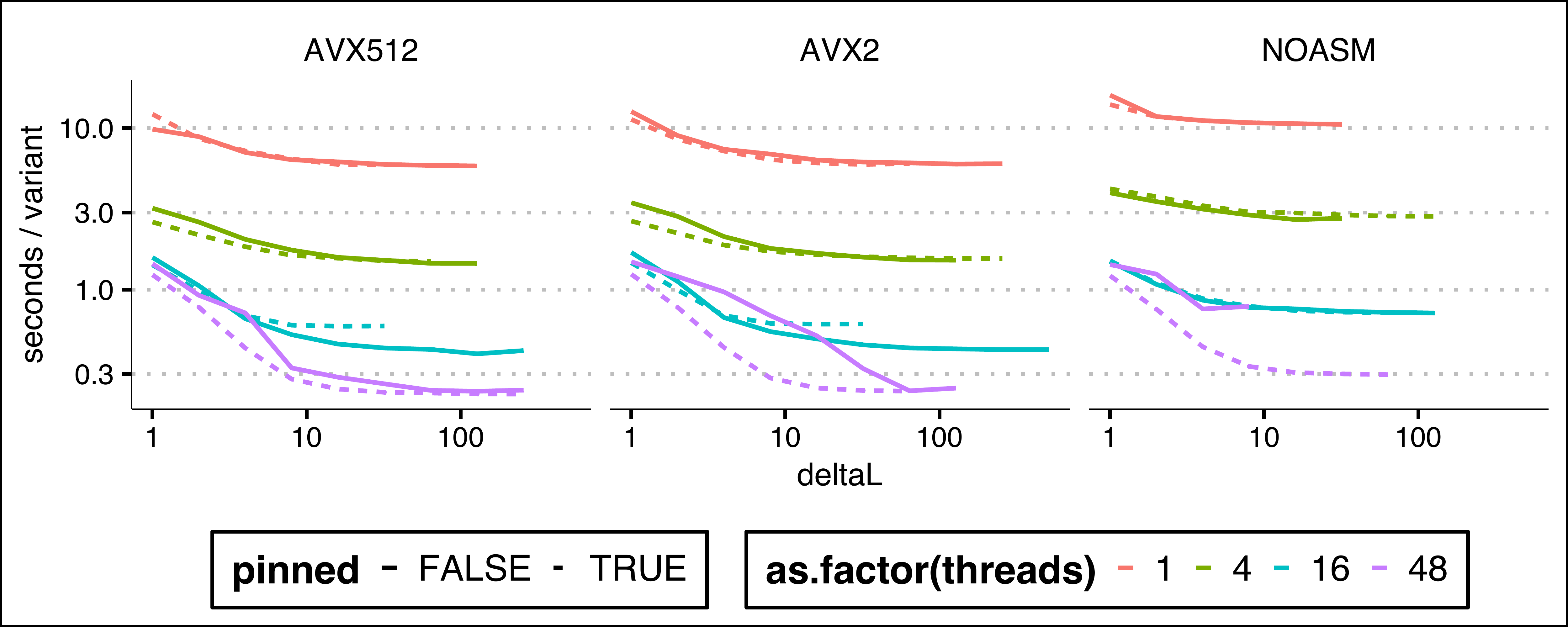}
    \caption{Log-log plot of seconds per variant performance ($y$-axis) of the forward algorithm on 100,000 haplotypes, against the number of variants propagated ($x$-axis). Each panel is a different instruction set (AVX-512/AVX2/none).  Line colour denotes number of CPU threads, whilst a dashed line indicates pinned threads and a solid line indicates no thread pinning.  In total, using AVX-512, 48 threads, and pinned threads, it takes less approximately 38 minutes to propagate a $100000 \times 100000$ forward table over 10,000 variants.}
  \label{fig:perfdeltaL2}
\end{figure}

Finally, \Cref{fig:perfdeltaL2} shows that in certain very large problem
settings \pkg{kalis}' ability to pin threads can make a substantial
difference. In this setting, AVX2 showed the greatest benefit from
eliminating context switching, ensuring that the cache is not
invalidated by threads migrating between cores. The lack of substantial
difference between AVX2 and AVX-512 here once thread pinning is employed
calls for some investigation, though this may be a result of thermal
throttling which is known to occur especially for AVX-512 heavy code.

These performance examples again highlight the importance of pilot
benchmark runs with different configurations of instruction set and
unroll settings before embarking on long compute runs to ensure the
greatest compute efficiency is achieved for a given problem and compute
architecture.

\hypertarget{sec:realdata}{%
\section{Real-data example: recent selection for lactase
persistence}\label{sec:realdata}}

\emph{LCT} is a gene on chromosome 2 that encodes lactase, the enzyme
responsible for the breakdown and digestion of lactose, the sugar
commonly found in milk. Ancestral humans had a regulatory `switch' on
chromosome 2 that stops lactase production after infancy when children
would be weaned off breast milk. Mutations that disrupt this switch
allow lactase production to persist into adulthood, conferring a
lifelong ability to extract energy from milk \citep{ingram2009lactose}.
Such mutations have arisen independently at least twice in human
history, in Europe and in East Africa, and are among the strongest
examples of recent positive natural selection in humans
\citep{ranciaro2014genetic, bersaglieri2004genetic}. These mutations
have been shown to spread across standard human population boundaries.
Using another implementation of the LS model, \citet{busby2017inferring}
estimated that a European haplotype conferring lactase persistence
became prevalent within the West African Fula population due to natural
selection sometime over the past two thousand years.

\begin{figure}
\centering
\includegraphics{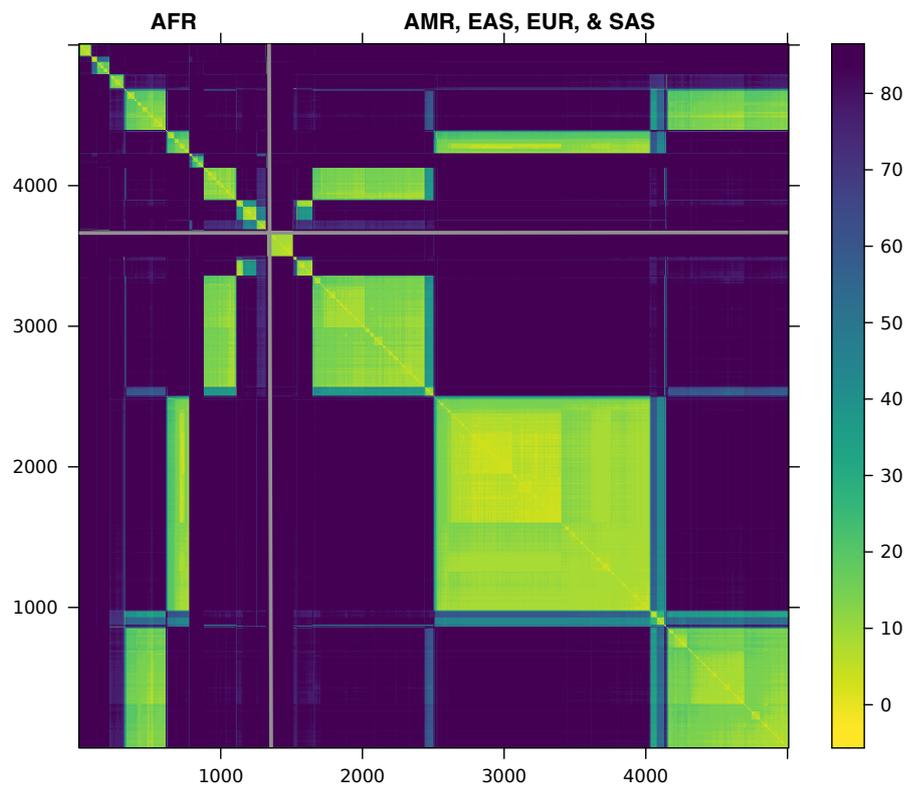}
\caption{\label{fig:lct}Distance matrix among 5008 haplotypes calculated
at \texttt{rs4988235}, upstream of \emph{LCT}. African haplotypes are
clustered in the upper left corner and separated by grey lines from
non-African haplotypes from the Americas (AMR), East Asia (EAS), Europe
(EUR), and SAS (South Asia). The scale on the right maps the colors to
distances.}
\end{figure}

Here we run \pkg{kalis} on 5008 haplotypes from the 1000 Genomes Phase 3
release to revisit the haplotype structure around \emph{LCT}; the
haplotypes are sampled from 26 sub-populations all over the world
\citep{10002015global}. \Cref{fig:lct} shows a clustered version of a
distance matrix, calculated as in \Cref{eq:distmat}, at a variant in the
regulatory region of \emph{LCT} (\texttt{rs4988235}). To see if we could
observe a pattern of gene-flow into or out of Africa similar to what was
observed by \citet{busby2017inferring}, we use average pairwise linkage
\citep{sokal1958statistical} to cluster the African haplotypes
separately from the non-African haplotypes. In \Cref{fig:lct}, distances
between African haplotypes are shown in the upper left corner;
non-African haplotypes, in the lower right corner.

Rather than 26 clusters reflecting the 26 sampled human populations, we
see that there are three very distinct lactase haplotypes that are
common both within and outside Africa. This suggests that these three
haplotypes, under strong positive selection pressure, recently spread
across population boundaries and presumably confer lactase persistence.
However, these three haplotypes are not the only structure we see: in
the upper left corner of the African (AFR) block we see some haplotypes
that are only found inside Africa; and in the non-African block, a
haplotype that is only found outside Africa. We can also see some
sub-structure within the clear haplotype blocks.

\hypertarget{sec:future}{%
\section{Future work}\label{sec:future}}

There are many avenues for future research in developing \pkg{kalis}. On
the model side, for example, allowing for different recombination rates
between sub-populations as done in fastPHASE \citep{scheet2006fast}
would be a natural extension.

On the computational side, ARM scalable vector extensions \citep{armsve}
represent an interesting new approach to SIMD instruction sets, where
the width of instructions need not be hard coded prior to compilation.
At present it is not widely available, but as this rolls out, it would
be natural to extend \pkg{kalis} to enable targeting this new
instruction set.

An important utility extension is expanding the file formats that
\pkg{kalis} can natively read via \texttt{CacheHaplotypes()}, to enable
simpler and more streamlined software pipelines when bioinformaticians
incorporate \pkg{kalis} into their workflows.

Finally, a future avenue of potential development is extension of
\pkg{kalis} to support GPU or tensor cards. Note that it was an explicit
design choice to initially target CPU SIMD extensions, since the vast
majority of University high performance computing clusters have a huge
amount of untapped compute power in this form, but often much more
limited availability of specialist extension cards. Therefore, by
pushing performance as extensively as possible via CPU only means, we
provide the greatest potential impact for end users. This does not
preclude future versions adding support for add-on compute cards.

\bibliography{kalis.bib}

\newpage
\appendix

\hypertarget{appendices}{%
\section*{Appendices}\label{appendices}}
\addcontentsline{toc}{section}{Appendices}

\hypertarget{mathematical-details-of-hmm-reformulation}{%
\section{Mathematical details of HMM
reformulation}\label{mathematical-details-of-hmm-reformulation}}

\hypertarget{apx:fwd}{%
\subsection{Rearrangement of the forward recursion}\label{apx:fwd}}

Starting with we \eqref{eq:raw_forward}, we have \begin{align}
  \tilde{\alpha}_{\cdot i}^{\ell} & \gets \theta^{\ell}_{\cdot i}  \left( \left(1-r^{\ell-1}\right) \tilde{\alpha}_{\cdot i}^{\ell-1} + r^{\ell-1} F_i^{\ell-1}  \pi_{\cdot i} \right) \nonumber \\
  \frac{\tilde{\alpha}_{\cdot i}^{\ell}}{ F_i^{\ell-1}} & \gets \theta^{\ell}_{\cdot i} \left( \left(1-r^{\ell-1}\right) \frac{1}{{ F_i^{\ell-1}}}\tilde{\alpha}_{\cdot i}^{\ell-1} + r^{\ell-1} \pi_{\cdot i} \right) \nonumber \\
  \frac{\tilde{\alpha}_{\cdot i}^{\ell}}{ F_i^{\ell-1}} & \gets \theta^{\ell}_{\cdot i} \left( \left(1-r^{\ell-1}\right) \frac{F_i^{\ell-2}}{{ F_i^{\ell-1}}} \frac{\tilde{\alpha}_{\cdot i}^{\ell-1}}{F_i^{\ell-2}} + r^{\ell-1} \pi_{\cdot i} \right) \nonumber \\
  \alpha_{\cdot i}^{\ell} & \gets \theta^{\ell}_{\cdot i} \left( \left(1-r^{\ell-1}\right) \frac{F_i^{\ell-2}}{{ F_i^{\ell-1}}} \alpha_{\cdot i}^{\ell-1} + r^{\ell-1} \pi_{\cdot i} \right) \nonumber
\end{align}

Since
\(\frac{F_i^{\ell-2}}{F_i^{\ell-1}} = \left( \frac{ \underset{j}{\sum} \tilde{\alpha}_{ji}^{\ell-1}}{ F_i^{\ell-2}}\right)^{-1} = \left( \underset{j}{\sum} \alpha_{ji}^{\ell-1} \right)^{-1}\),
we arrive at \eqref{eq:fwd1}.

\hypertarget{apx:bck}{%
\subsection{Rearrangement of the backward recursion}\label{apx:bck}}

Starting with \eqref{eq:raw_backward} we have

\begin{align}
  \tilde{\beta}_{\cdot i}^{\ell} &\gets \left( 1- r^{\ell}\right) \tilde{\beta}_{i\cdot}^{\ell+1} \theta^{\ell+1}_{\cdot i} + r^{\ell} G^\ell \nonumber \\
  \frac{\tilde{\beta}_{\cdot i}^{\ell}}{G^{\ell}} &\gets \left( 1- r^{\ell}\right) \frac{1}{G^{\ell}} \tilde{\beta}_{\cdot i}^{\ell+1} \theta^{\ell+1}_{\cdot i} + r^\ell \nonumber \\
  \frac{\tilde{\beta}_{\cdot i}^{\ell}}{ G^{\ell}} &\gets \left( 1- r^{\ell}\right) \frac{G^{\ell+1}}{G^{\ell}} \frac{\tilde{\beta}_{\cdot i}^{\ell+1}}{G^{\ell+1}} \theta^{\ell+1}_{\cdot i} + r^\ell \nonumber \\
  \beta_{\cdot i}^{\ell} &\gets \left( 1- r^{\ell}\right) \frac{G^{\ell+1}}{ G^{\ell}} \beta_{\cdot i}^{\ell+1} \theta^{\ell+1}_{\cdot i} + r^\ell \nonumber \\
  \label{backward_almost_rearranged}
\end{align}

Since
\(\frac{G^{\ell+1}}{ G^{\ell}} = \left( \frac{\underset{j}{\sum} \tilde{\beta}_{ji}^{\ell+1}\theta_{ji}^{\ell+1} \pi_{ji} }{G_i^{\ell+1}}\right)^{-1} = \left(\underset{j}{\sum} \beta_{ji}^{\ell+1}\theta_{ji}^{\ell+1} \pi_{ji}\right)^{-1}\),
we arrive at \eqref{eq:bck1}.

\subsection[Numerical considerations: avoiding NaN]{Numerical
considerations: avoiding \texttt{NaN}}\label{apx:nan}

The forward and backward recursions as written in \eqref{eq:fwd1} and
\eqref{eq:bck1} are susceptible to three main categories of numerical
instability: element underflow, total underflow, and overflow. By
element underflow, we refer to the situation where a subset of elements
in \(\alpha_{\cdot i}^\ell\) or \(\beta_{\cdot i}^\ell\) underflow to
zero for a given \(\ell\). This becomes more likely to occur if there
are zero (or near zero) entries in \(\rho\), \(\mu\), or \(\Pi\).
Element underflow effectively reinitializes the recursion for the donor
haplotypes corresponding to the elements where the underflow occurs,
causing the HMM to lose track of how similar those donor haplotypes are
to the recipient haplotype leading up to the variant where the underflow
occurs. While element underflow results in a loss of information about
the relative likelihood of the recipient copying from genetically
distant donors, the relative likelihood of copying similar donors is
retained. Element underflow will not cause either recursion to fail.

Underflow can cause catastrophic failure if it causes either
\(\underset{j}{\sum} \alpha_{ji}^{\ell-1}\) or
\(\underset{j}{\sum} \beta_{ji}^{\ell+1}\theta_{ji}^{\ell+1} \pi_{ji}\)
to evaluate to zero at a given \(\ell\), which we refer to as total
underflow. In these cases, entire columns of \texttt{fwd\$alpha} or
\texttt{bck\$beta} will be \texttt{NaN} (except for the diagonal which
is always 0). While this means that the user cannot continue the
recursion with the current set of haplotypes and parameters, total
underflow easy to catch since the \texttt{NaN}s will usually break
downstream pipelines.

Internally, \pkg{kalis} takes several measures to help reduce the risk
of total underflow. For example, \pkg{kalis} calculates
\(\theta_{\cdot i}^{\ell}\) as \begin{equation}
  \theta_{\cdot i}^{\ell} = \left(1 - H_{\cdot i}^{\ell}\right)*\left(1-2\mu\right) + \mu .
\end{equation}

Note, that even for \(\mu\) well below machine precision, this sets
\(\theta_{j i}^{\ell}\) to \(\mu\) for haplotypes \(j,i\) that mismatch
at \(\ell\) rather than setting \(\theta_{j i}^{\ell}\) to 0. This in
turn helps prevent \(\theta_{\cdot i}^{\ell}\) from being set to zero,
resulting in total underflow. Despite precaution taken in \pkg{kalis},
we recommend that users remove private mutations that appear on only one
haplotype (singletons) before loading haplotypes into the \pkg{kalis}
cache. These private mutations tend not to be informative about the
relationships between haplotypes and removing them can help prevent
total underflow, especially at variants with small \(\mu\).

In addition to removing singletons, users may consider avoiding
parameter choices that place many zero (or near zero) entries in
\(\rho\), \(\mu\), or \(\Pi\) to help prevent or remedy total underflow.
Setting some entries of \(\rho\) zero, for a non-recombining segment of
genome, or \(\mu\) to zero, for very important variants that all
potential donors must share with a recipient, is often biologically
meaningful and should be safe in most applications. However, setting
entries of \(\Pi\) to zero requires slightly more caution because this
can easily cause total underflow if the prior copying probabilities
strongly conflict with the observed haplotype similarity over a genomic
interval (prior-likelihood mismatch). Furthermore, zero (or very near
zero) prior copying probabilities can cause
\(\underset{j}{\sum} \beta_{ji}^{\ell+1}\theta_{ji}^{\ell+1} \pi_{ji}\)
to evaluate to \texttt{Inf} which we refer to as total overflow when
there is prior-likelihood mismatch. Like total underflow, total overflow
is easily detectable since it generate \texttt{NaN}s on the next
backward iteration. The forward recursion is not susceptible to total
overflow because every \(\alpha_{ji}^\ell\) is less than or equal to 2
\footnote{In a different implementation, one may be tempted to divide
  both sides of the forward recursion and backward recursion by
  \(\rho\). This would help prevent underflow by allowing \(\alpha\) to
  make full use of the range of double precision numbers and have the
  added benefit of slightly increased performance (by eliminating a
  multiply AVX instruction). However, rescaling by \(\rho\) makes it
  impossible to tackle problems where \(\rho^\ell\) is zero for some
  \(\ell\). Even if we restrict \(\rho^\ell > 0\), the rescaling makes
  the forward recursion susceptible to total overflow since it raises
  the upper bound on \(\alpha\) from 2 to
  \(1 + \left( \underset{\ell}{\min} \rho^\ell \right)^{-1}\).
  Similarly, it makes the backward recursion more susceptible to
  overflow by raising the upper bound on \(beta\) from
  \(1 + \left( \underset{j,i}{\min} \pi_{ji} \right)^{-1}\) to
  \(1 + \left( \underset{\ell,j,i}{\min} \rho^\ell \pi_{ji} \right)^{-1}\).}.
Using prior copying probabilities that reflect the observed haplotypic
similarity or simply resorting to uniform prior copying probabilities
for \(\Pi\) (the default) should be safe in most applications.

\hypertarget{installation-help}{%
\section{Installation help}\label{installation-help}}

There are two key features that can only be set at package compile time:
the SIMD instruction set to target, and how deeply to unroll the
innermost loop in the \pkg{kalis} core. This appendix explains how to
set these options at compile time.

\hypertarget{apx:compileis}{%
\subsection{Manually controlling instruction set}\label{apx:compileis}}

When compiling \pkg{kalis} it will by default attempt to auto-detect the
best available SIMD instruction set to use. However, if this
auto-detection fails, or if you wish to force the use of an inferior (or
no) SIMD instructions then you can use a compiler flag to manually
direct \pkg{kalis}.

The supported flags are:

\begin{CodeChunk}
\begin{longtable}{ll}
\toprule
\textbf{Flag} & \textbf{Instruction sets used} \\ 
\midrule
\texttt{NOASM} & Forces pure C, with no special instruction set intrinsics used \\ 
\texttt{AVX2} & Enables intrinsics for AVX2, AVX, SSE4.1, SSE2, FMA and BMI2 \\ 
\texttt{AVX512} & Enables intrinsics for AVX-512, AVX2, SSE2 and BMI2 \\ 
\texttt{NEON} & Enables intrinsics for ARM NEON and NEON FMA \\ 
\bottomrule
\end{longtable}
\end{CodeChunk}

They are used by setting the configure variable \texttt{FLAG=1}, where
\texttt{FLAG} is one of the options in the above table, just as shown in
\Cref{sec:installing}. For example, to force the use of no special
instruction set whilst still compiling against the native architecture,
one would compile with:

\begin{verbatim}
remotes::install_github("louisaslett/kalis",
  configure.vars =
    c(kalis = "PKG_CFLAGS='-march=native -mtune=native -O3' NOASM=1"))
\end{verbatim}

or pulling the source from CRAN,

\begin{verbatim}
install.packages("kalis", type = "source",
  configure.vars =
    c(kalis = "PKG_CFLAGS='-march=native -mtune=native -O3' NOASM=1"))
\end{verbatim}

Note that the most restrictive instruction set is all that need be
defined: that is, the availability of AVX2 implies the availability of
the other listed instruction sets, and likewise for AVX-512 and NEON.
Hence, there is no need to verify the availability of other instruction
sets.

On a Mac, you can check if you have any of these instruction sets by
running the following at a Terminal, the lines ending \texttt{1}
indicating support (missing lines or those ending \texttt{0} indicate no
support):

\begin{verbatim}
sysctl -a | egrep "^hw.optional.(avx2|avx5|neon)+"
\end{verbatim}

\hypertarget{apx:unroll}{%
\subsection{Core loop unrolling}\label{apx:unroll}}

Loop unrolling can improve performance for critical deeply nested loops.
This occurs either because for a sufficiently optimised loop, the loop
increment count comprises a substantial proportion of the computation of
each iteration, or because by unrolling the compiler (and CPU at run
time) can reason about between iteration dependency better leading to
better instruction ordering and potential instruction level parallelism.

By default \pkg{kalis} will unroll loops to depth 4, which we have
tested to be a reasonable default for many machines and problem sizes.
However, the optimal value will vary both by the particulars of a
CPU/memory architecture and by problem size (in \(N\)). Therefore, if
\pkg{kalis} represents a performance critical section of your code base,
we recommend running real benchmarks for a variety of unroll depths on a
problem of your target size on the machine you will deploy to.

In order to set the unroll depth, you need to pass the \texttt{UNROLL}
configure variable. Note that only powers of 2 are supported.

For example, to double the default unroll depth to 8, use the following
at compile time:

\begin{verbatim}
remotes::install_github("louisaslett/kalis",
  configure.vars =
    c(kalis = "PKG_CFLAGS='-march=native -mtune=native -O3' UNROLL=8"))
\end{verbatim}

or pulling the source from CRAN,

\begin{verbatim}
install.packages("kalis", type = "source",
  configure.vars =
    c(kalis = "PKG_CFLAGS='-march=native -mtune=native -O3' UNROLL=8"))
\end{verbatim}

Note that the unroll flag and the target SIMD instruction set flags of
\Cref{apx:compileis} can be set together with a space separator between
them.

\hypertarget{apx:hdf5}{%
\section{HDF5 file format}\label{apx:hdf5}}

HDF5 \citep{hdf5} is a format designed to handle large quantities of
data in an efficient manner. \pkg{kalis} supports loading data from HDF5
in the format specified here, with the option to depend on either CRAN
package \pkg{hdf5r} \citep{hdf5r} or Bioconductor \citep{bioc} package
\pkg{rhdf5} \citep{rhdf5}.

For HDF5 files, \pkg{kalis} expects a 2-dimensional object named
\texttt{/haps} at the root level of the HDF5 file. Haplotypes should be
stored in the slowest changing dimension as defined in the HDF5
specification (\textbf{note:} different languages treat this as rows or
columns: it is `row-wise' in the \proglang{C} standard specification, or
`column-wise' in the \pkg{rhdf5} specification). If the haplotypes are
stored in the other dimension then simply supply the argument
\texttt{transpose\ =\ TRUE} when calling \texttt{CacheHaplotypes()},
although this may incur a small penalty in the time it takes to load
into cache. If you are unsure of the convention of the language in which
you created the HDF5 file, then the simplest approach is to simply load
the data with \texttt{CacheHaplotypes()} specifying only the HDF5 file
name and then confirm that the number of haplotypes and their length
have not been exchanged in the diagnostic output which \pkg{kalis}
prints.

The format also allows named IDs for both haplotypes and variants in the
1-dimensional objects \texttt{/hap.ids} and \texttt{/loci.ids}, also in
the root level of the HDF5 file.

\subsection[Working with this format in R]{Working with this format in
\proglang{R}}\label{apx:writehdf5}

Since the format described above is not standard, \pkg{kalis} provides
two utility functions for working with it. \texttt{WriteHaplotypes()}
enables using \proglang{R} to create the requisite format from a
standard matrix, and \texttt{ReadHaplotypes()} allows reading into
\proglang{R} (rather than the internal \pkg{kalis} cache) from this
format.

Assume that you have imported your haplotype data into the \proglang{R}
variable \texttt{myhaps}, with variants in rows and haplotypes in
columns (so that \texttt{myhaps} is an \(L \times N\) matrix consisting
of only \texttt{0} or \texttt{1} entries). Then to write this out to the
HDF5 file \texttt{\textasciitilde{}/myhaps.h5},

\begin{CodeChunk}
\begin{CodeInput}
R> WriteHaplotypes("~/myhaps.h5", myhaps)
\end{CodeInput}
\end{CodeChunk}

You may additionally wish to store the names of haplotypes or variants
alongside the haplotype matrix for self documentation purposes. Assume
you have defined \texttt{hapnm}, a character vector of length \(N\), and
\texttt{varnm}, a character vector of length \(L\), then you may instead
call,

\begin{CodeChunk}
\begin{CodeInput}
R> WriteHaplotypes("~/myhaps.h5", myhaps, hap.ids = hapnm, loci.ids = varnm)
\end{CodeInput}
\end{CodeChunk}

You can verify the content by reading this file back into R directly as:

\begin{CodeChunk}
\begin{CodeInput}
R> myhaps_file <- ReadHaplotypes("~/myhaps.h5")
\end{CodeInput}
\end{CodeChunk}

You should find \texttt{all(myhaps\ ==\ myhaps\_file)} is \texttt{TRUE}.
For large problems, you may want to read just one named variant. For
example, imagine you had saved the variant names using the second
\texttt{WriteHaplotypes()} function call above, and that one of those
variant names was \texttt{"rs234"} (genetic variant associated with
lactase persistence), then you can read all haplotypes for this variant
with,

\begin{CodeChunk}
\begin{CodeInput}
R> lactase_haps <- ReadHaplotypes("~/myhaps.h5", loci.ids = "rs234")
\end{CodeInput}
\end{CodeChunk}

Once you are happy the HDF5 file is setup correctly, you can restart
your \proglang{R} session to ensure all memory is cleared and then load
your haplotype data directly to the optimised \pkg{kalis} cache by
passing the file name,

\begin{CodeChunk}
\begin{CodeInput}
R> CacheHaplotypes("~/myhaps.h5")
\end{CodeInput}
\end{CodeChunk}

\end{document}